\documentclass[structabstract]{aa}
\usepackage{txfonts}
\usepackage{natbib}
\usepackage{graphicx}
\usepackage{aalongtable}
\bibpunct{(}{)}{;}{a}{}{,}

\begin{document}

\title{Gravito-inertial and pressure modes detected in the B3\,IV
  CoRoT target HD\,43317 \thanks{The CoRoT space mission was developed and is
    operated by the French space agency CNES, with participation of ESA's RSSD
    and Science Programmes, Austria, Belgium, Brazil, Germany, and
    Spain.}\fnmsep\thanks{Based on data gathered with \textsc{Harps} installed
    on the 3.6 meter ESO telescope (ESO Large Programme 182.D-0356) at La Silla,
    Chile.}\fnmsep\thanks{Table\,\ref{frequtable} is only available in electronic form
    at the CDS via anonymous ftp to cdsarc.u-strasbg.fr (130.79.128.5)
    or via http://cdsweb.u-strasbg.fr/cgi-bin/qcat?J/A+A/}}

\author{P.~I.~P\'{a}pics\inst{\ref{inst1}}
\and M.~Briquet\inst{\ref{inst2}}\thanks{F.R.S.-FNRS Postdoctoral Researcher, Belgium.}
\and A.~Baglin\inst{\ref{inst3}} 
\and E.~Poretti\inst{\ref{inst4}}
\and C.~Aerts\inst{\ref{inst1},\ref{inst5}}
\and P.~Degroote\inst{\ref{inst1}}
\and A.~Tkatchenko\inst{\ref{inst1}}
\and T.~Morel\inst{\ref{inst2}}
\and W.~Zima\inst{\ref{inst1}}
\and E.~Niemczura\inst{\ref{inst6}}
\and M.~Rainer\inst{\ref{inst4}}
\and M.~Hareter\inst{\ref{inst7}}
\and F.~Baudin\inst{\ref{inst8}}
\and C.~Catala\inst{\ref{inst9}}
\and E.~Michel\inst{\ref{inst9}}
\and R.~Samadi\inst{\ref{inst9}}
\and M.~Auvergne\inst{\ref{inst9}}}

\institute{Instituut voor Sterrenkunde, K.U.Leuven, Celestijnenlaan 200D, B-3001 Leuven, Belgium \email{Peter.Papics@ster.kuleuven.be}\label{inst1}
\and Institut d'Astrophysique et de G\'{e}ophysique, Universit\'{e} de Li\`{e}ge, All\'{e}e du 6 Ao\^{u}t 17, B\^{a}t B5c, 4000 Li\`{e}ge, Belgium\label{inst2}
\and Laboratoire AIM, CEA/DSM-CNRS-Universit\'{e} Paris Diderot; CEA, IRFU, SAp, centre de Saclay, F-91191, Gif-sur-Yvette, France\label{inst3}
\and INAF - Osservatorio Astronomico di Brera, via E.~Bianchi 46, 23807 Merate (LC), Italy\label{inst4}
\and Department of Astrophysics, IMAPP, University of Nijmegen, PO Box 9010, 6500 GL Nijmegen, The Netherlands\label{inst5}
\and Instytut Astronomiczny, Uniwersytet Wroclawski, Kopernika 11, 51-622, Wroclaw, Poland\label{inst6}
\and Institut f\"{u}r Astronomie, Universit\"{a}t Wien, T\"{u}rkenschanzstrasse 17, A-1180 Vienna, Austria\label{inst7}
\and Institut d'Astrophysique Spatiale, CNRS/Univ. Paris-Sud, B\^{a}t. 121, F-91405, Orsay Cedex, France\label{inst8}
\and LESIA, UMR8109, Universit\'{e} Pierre et Marie Curie, Universit\'{e} Denis Diderot, Observatoire de Paris, 92195 Meudon Cedex, France\label{inst9}}

\date{Received 11 January 2012 / Accepted 21 March 2012}

\abstract{OB stars are important building blocks of the Universe, but we have
  only a limited sample of them well understood enough from an
  asteroseismological point of view to provide feedback on the current evolutionary
  models. Our study adds one special case to this sample, with more
  observational constraints than for most of these stars.}{Our goal is to analyse
  and interpret the pulsational behaviour of the B3\,IV star HD\,43317 using the
  CoRoT light curve along with the ground-based spectroscopy gathered by the
  \textsc{Harps} instrument. This way we continue our efforts to map the
  $\beta$\,Cep and SPB instability strips.}{We used different techniques to
  reveal the abundances and fundamental stellar parameters from the
  newly-obtained high-resolution spectra. We used various time-series analysis
  tools to explore the nature of variations present in the light curve. We
  calculated the moments and used the pixel-by-pixel method to look for line
  profile variations in the high-resolution spectra.}{We find that HD\,43317 is
  a single fast rotator ($v_{\mathrm{rot}}\approx50\%\,v_{\mathrm{crit}}$) and
  hybrid SPB/$\beta$\,Cep-type pulsator with Solar metal
  abundances. We interpret the variations in photometry and spectroscopy as a
  result of rotational modulation connected to surface inhomogeneities, combined
  with the presence of both $g$ and $p$ mode pulsations. We detect a series of
  ten consecutive frequencies with an almost constant period spacing of 6339\,s
  as well as a second shorter sequence consisting of seven frequencies with a
  spacing of 6380\,s. The dominant frequencies fall in the regime of
  gravito-inertial modes.}{}

\keywords{Asteroseismology - 
Stars: variables: general - 
Stars: abundances -
Stars: oscillations - 
Stars: individual: HD\,43317 - 
Stars: rotation}

\titlerunning{Gravito-inertial and pressure modes in the B3\,IV star HD\,43317}
\maketitle


\section{Introduction}\label{intro}

Asteroseismology is a tool which can be regarded as an  
\textit{astrophysical stethoscope}. The detection and analysis of the effects of 
sound waves propagating through stellar interiors is one of the few ways in which we can 
deduce information about the insides of stars. 
We use light and radial velocity variations from pulsating stars along with
theoretical models -- continuously refined by the observations -- to deduce
information from below their surface. With the launch of space-based
instruments such as the CoRoT \citep[Convection Rotation and planetary
Transits,][]{2009A&A...506..411A} and {\it Kepler\/} \citep{2010PASP..122..131G}
satellites, the availability of virtually
uninterrupted photometric data sets at micro-magnitude precision has enabled us to
sharpen our view, and opened up a new era in asteroseismology.

In this paper we are concerned with massive stars which are important building
blocks in the chemistry of our Universe, enriching the interstellar matter --
when they end their lives in a supernova explosion -- with heavy elements they
produced during their lifetimes. From an asteroseismic point of view, they are
relatively easy to interpret. Stellar lifetime is strongly influenced by several
internal mixing processes, such as those due to core overshooting and rotation,
which are the major uncertainties in the present-day models.  With the precision
of data delivered by currently available space instruments, we aim to deduce a
quantitative seismic estimate of the overshooting and internal rotation in B
stars to calibrate stellar evolution models in the upper part of the
Hertzsprung-Russel diagram (HRD) from a carefully selected sample of suitable
pulsators.

Slowly rotating pulsating OB stars situated in the upper part of the HRD have a
relatively simple frequency content. In general, the more massive and hot
$\beta$\,Cep stars oscillate in low-order $p$ (and $g$) modes with typical
periods between about 3 and 8 hours, while the less massive and
cooler Slowly Pulsating B (SPB) stars oscillate in high-order $g$ modes with
periods between 0.8 and 3 days
\citep{2010aste.book.....A}. This simple picture gets more complicated by considering,
for example, the presence of rotation. On the other hand, slow rotation can help mode
identification by detecting multiplets of different modes. Given that these penetrate to
different depths in the star, rotationally split multiplets offer the potential
to infer the internal rotational profile
\citep[e.g.,][]{1999ApJ...516..349K,2003Sci...300.1926A}.

Based on CoRoT data seismic inferences on the interior structure of slowly
rotating pulsating OB stars has been derived. While the power of present-day state-of-the-art
modelling has been demonstrated for the B1.5\,II-III $\beta\,$Cep star HD\,180642
\citep{2011A&A...534A..98A}, unexpected stochastic modes with constant frequency
spacings were detected in the O8.5\,V star HD\,46149
\citep{2010A&A...519A..38D}. On the other hand, heat-driven $\beta\,$Cep-like
modes were detected but not predicted to be excited in the O9\,V star HD\,46202
\citep{2011A&A...527A.112B} and vice-versa in the B0.5\,IV star HD\,51756
\citep{2011A&A...528A.123P}.  \citet{2010Natur.464..259D} discovered an almost
constant period spacing of the high-order $g$ modes in the frequency spectrum of
the B3\,V hybrid SPB/$\beta$\,Cep star HD\,50230, serving as a viable diagnostic
to deduce the overshooting parameter for B pulsators. Moreover, the small
periodic deviations from this spacing were used to model the diffusive mixing
process around the core.

The B3\,IV star HD\,43317 was selected as our target in this context, as part of our
attempts to increase the number of well-studied B stars. We were especially
looking for a star with nearly identical parameters as those of HD\,50230, but
brighter, to see if it has similar behaviour of its $g$ modes and to deduce
the implications of a higher rotation velocity. In parallel with the monitoring
done by CoRoT's asteroseismology CCD, we gathered ground-based spectroscopy to
give as many observational constraints as possible, thus providing the
foundations for stellar modelling.


\section{Fundamental parameters}\label{fundparameters}

\subsection{Prior to our study}\label{prior}

HD\,43317 is a bright field-star in the constellation of Orion. It was first
mentioned in proper motion studies \citep[e.g.,][]{1956AJ.....61...90M} and
survey catalogues of early-type stars
\citep[e.g.,][]{1967JO.....50..237B}. Further investigations include
interstellar absorption and extinction measurements
\citep{1971ApJ...166..543W,1985ApJS...59..397S} and Galactic kinematic studies
\citep[e.g.,][]{1985A&AS...60...99W}, which are all common for B type
stars. This is the first time that HD\,43317 is investigated from an
asteroseismic point of view.

There is good agreement on both the spectral type and luminosity class: B3\,IV
\citep {1968ApJS...17..371L}, B2\,V \citep{1968PASP...80..197G}, and B3\,IV/V
\citep{1999mctd.book.....H}. These classifications are all based on visual
comparison of objective-prism plates to standard stars from a spectral
atlas. There is one classification of B2.5\,V using ultraviolet line features of
S2/68 spectra \citep{1976A&AS...26..241C}. Effective temperature estimates (based on the spectral type and different calibration tables) of
17\,100\,K and 17\,900\,K are given by \citet{2003AJ....125..359W} and
\citet{2010AN....331..349H}. \citeauthor{2010AN....331..349H} also give an estimated mass of
$6.05\pm0.69\mathcal{M}_{\odot}$.  Due to lack of high-quality spectroscopy,
\citet{2010A&A...515A..74L} could not include this star in their large sample of
B stars in the CoRoT field-of-view to deduce the fundamental parameters and
abundances with better precision.

Quoted estimates of the projected rotational velocity $v\sin i$ are
$110\pm10\mathrm{\,km\,s}^{-1}$ \citep{1982ApJ...252..322W} and $130\pm16\mathrm{\,km\,s}^{-1}$
\citep[]{2002ApJ...573..359A}. Distance measurements of $369^{+73}_{-53}$ pc
\citep[from a parallax of $2.71\pm0.45\,\mathrm{mas}$,][]{2007A&A...474..653V},
640\,pc \citep{1985ApJS...59..397S}, 446\,pc \citep{1985ApJ...294..599S}, and
598\,pc \citep{2000AJ....119..923H} are available from different methods.
Photometry led to $V_\mathrm{mag}=6.64$, $(B-V)=-0.17$, $(U-B)=-0.66$, and
$H_{\beta}=2.665$ \citep{1971AJ.....76.1048C}; the data available for many more
filters will be used in Sect.\,\ref{SED}. Colour excess estimates of $E(B-V)$ are between 
$0.03$ \citep{1985ApJS...59..397S} and $0.10$ \citep{2000AJ....119..923H}. From these values 
we derived an absolute magnitude estimate of $M_\mathrm{V}=-1.80\pm0.91$ which is consistent with the 
spectral type and luminosity class of the star \citep[based on the tables provided by][]{1981Ap&SS..80..353S}.

\subsection{SED-fitting}\label{SED}

\begin{figure}
\resizebox{\hsize}{!}{\includegraphics{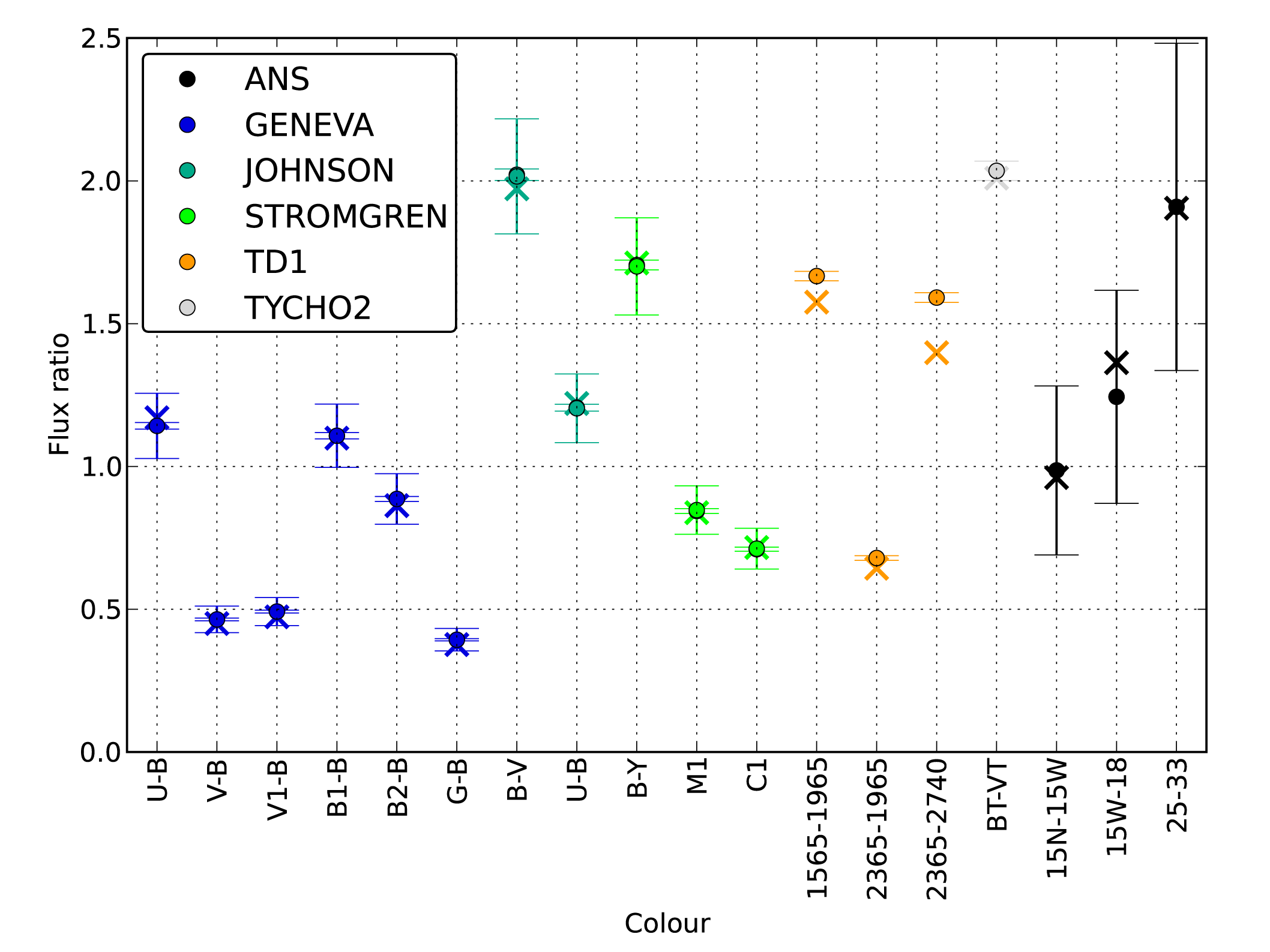}} 
\caption{Fitted flux ratios. Different marker colours denote
  different photometric systems, filter names are given on the x-axis. 'X' symbols are synthetic (model)
  colours, circles are observed ones.}
\label{photcolours}
\end{figure}

In an attempt to give better estimates of the fundamental parameters of
HD\,43317, the parameters were determined from multicolour photometry following
the method described in \citet{degroote2011}. We used UV \citep[ANS,
TD1,][]{cat_ans,calib_ans,cat_td1}, optical \citep[Geneva, Johnson, Hipparcos,
Tycho, Stromgren,][]{cat_geneva,calib_geneva,cat_ubv,cat_tycho,cat_uvby}, and
near-infrared \citep[2MASS, Wise,][]{2mass,calib_wise,cat_wise} photometry.  No
infrared excess was detected up to 20\,$\mu$m in the Akari \citep{akari} and
Wise bands. The photometry was fitted to LTE \citep{castelli2003} and non-LTE
\citep{2007ApJS..169...83L} model atmospheres, using a method similar to the
infrared-flux method by \citet{irfm}, i.e., by using absolute calibrated values
in the infrared and colours at shorter wavelengths. The best fit using the NLTE
TLUSTY models \citep{2007ApJS..169...83L} is shown in Fig.\,\ref{photcolours}, while the results are listed
in Table\,\ref{fundparphot}.

\begin{table}
  \caption{Stellar parameters derived from multicolour photometry using different model atmospheres.}
\label{fundparphot}
\centering
\renewcommand{\arraystretch}{1.25}
\begin{tabular}{l c c}
\hline\hline
Model									&Kurucz							&TLUSTY\\
\hline
$T_\mathrm{eff}\,(\mathrm{K})$ 			&$17\,200^{+900}_{-700}$			&$17\,350^{+950}_{-650}$\\
$\log g\,(\mathrm{cgs})$					&$4.75^{+0.25}_{-1.15}$			&$4.75^{+0.00}_{-0.75}$\\
$\log Z/Z_{\odot}$						&$-2^{+1}_{-0.5}$				&$-0.8^{+0.5}_{-0.2}$\\
$E(B-V)\,(\mathrm{mag})$					&$0^{+0.02}_{-0.00}$				&$0^{+0.03}_{-0.00}$\\
$\theta\,(\mathrm{mas})$					&$0.095^{+0.003}_{-0.003}$		&$0.095^{+0.002}_{-0.003}$\\
\hline
\end{tabular}
\tablefoot{Error ranges are set around the best fit value corresponding to the $90\%$ confidence interval based on the $\chi^2$ values. $Z_{\odot}=0.02\,\mathrm{dex}$.}
\end{table}

We will use recently obtained high-resolution spectra to derive the fundamental
parameters along with the abundances in Sect.\,\ref{harpsspectroscopy}. This
will be especially useful for the surface gravity value, as the
photometric method is fairly insensitive to it, which is the reason why the best
fit values are larger than the typical ZAMS value of $\sim4.5$, and for the
metallicity, which is also poorly constrained from photometry.


\section{The CoRoT light curve}

\subsection{Frequency analysis}\label{frequanal}

\begin{figure*}
\resizebox{\hsize}{!}{\includegraphics{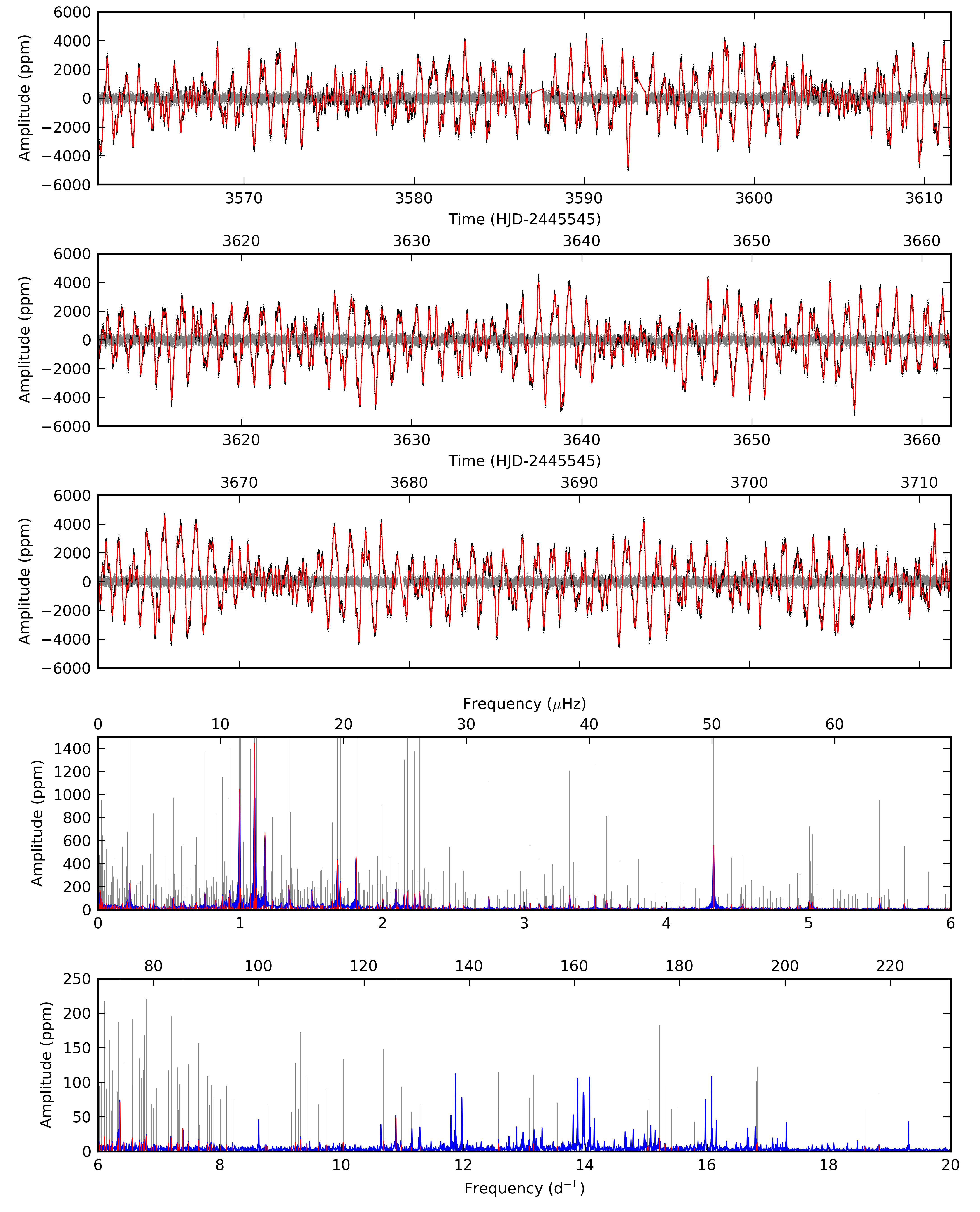}} 
\caption{(\textit{upper three panels}) The reduced CoRoT light curve (black dots, brightest at the top) and residuals (gray dots) after prewhitening with a model (red solid line) constructed using the set of 424 frequencies listed in Table\,\ref{frequtable} -- see text for further explanation. (\textit{lower two panels}) The Scargle periodogram of the full CoRoT light curve (blue solid line) showing the 424 frequencies (red vertical lines). For better visibility, the red lines are repeated in gray in the background, after multiplying their amplitude with a factor ten. Aliases of the strongest peaks around $\sim1\,\mathrm{d}^{-1}$ due to the satellite orbital motion can be seen around 12, 14 and 16\,$\mathrm{d}^{-1}$.}
\label{corotlcurvefourier}
\end{figure*}

HD\,43317 was observed by the CoRoT satellite as a primary target during the
LRa03 long run of the mission as part of the asteroseismology programme, while
monitoring a field in the direction of the anticentre of the Milky Way from
HJD 2455106.410969 for 150.41 days (covering five months from 1 October 2009 to
1 March 2010). All flagged observations were removed from the light curve,
leaving us with 359\,704 measurements (resulting in a duty cycle of almost
90\%). It is necessary to delete the flagged CoRoT data points because of the 
passage of the satellite through the South Atlantic anomaly and some other 
low Earth orbit perturbations \citep[see][]{2009A&A...506..411A}. This introduces small 
gaps in the light curve in phase with the orbital period, which produces a spectral window 
with aliases at the orbital frequency ($6184\,\mathrm{s}$, $13.97\,\mathrm{d}^{-1}$, $161.7\,\mu\mathrm{Hz}$) 
and its harmonics, at the frequency related to day/night variations ($2\,\mathrm{d}^{-1}$, $23.15\,\mu\mathrm{Hz}$), 
and combinations of them. The most prominent aliases appear at frequencies 
$f_{j}\pm n \times 13.97\pm1\pm2\,\mathrm{d}^{-1}$ , where $f_{j}$ is an intrinsic 
frequency, and $n\in\mathbb{N}^{\mathrm{*}}$ \citep[see, e.g.,][]{2009A&A...506..133G}. 
These aliases are removed during the prewhitening process. 
We looked for trends which have a well known instrumental origin
\citep{2009A&A...506..411A}, and modelled them with various prescriptions. In
order to correct the light curve for these instrumental effects, we divided it
by a fitted linear polynomial. As after this initial step, there were no clear
jumps or trends visible anymore, we used the resulting dataset (see
Fig.\,\ref{corotlcurvefourier}) in our analysis.

\subsubsection{Common approach}

The way one defines significance affects the final number of accepted
frequencies in a Fourier-analysis. In the commonly used approach of
\citet{1993A&A...271..482B}, one measures the signal-to-noise (S/N) level as the
ratio of a given peak to the average residual noise in a given window around the
peak or in a peak-free region, before (or after) prewhitening. This approach
worked fine for ground-based photometry, with relatively few significant peaks
in the data sets.  With space-based instruments like CoRoT and {\it Kepler\/}
delivering almost uninterrupted data with a precision typically two orders of
magnitude better than ground-based photometry, forests of peaks emerge above the
noise, and using the classical criteria is no longer obvious.

A simple model which uses a limited number of sine functions to describe the
pulsations of a star is a good approach when it comes to tens of frequencies,
but when hundreds or even thousands of them occur, each leading to low
variance reduction \citep[see, e.g.,][]{2009A&A...506...85P,
  2011A&A...525A..23C}, we prefer a more conservative approach and accept less
but more secure frequencies. Should any irregularities or previously undetected
features arise thanks to the higher precision, we must be able to detect them
without going down to the noise level.

We first performed an iterative prewhitening procedure whose description is
already given by \citet{2009A&A...506..111D} and thus omitted here. This
resulted in a list of amplitudes ($A_j$), frequencies ($f_j$), and phases
($\theta_j$), by which the light curve can be modelled via $n_f$ frequencies in
the well-known form of \[F(t_i)=c+\sum_{j=1}^{n_f}A_j\sin[2\pi(f_j t_i +
\theta_j)].\] The prewhitening procedure was stopped after removing 1353
frequencies (adopting a $p$ value of $p = 0.001$ in hypothesis testing). 
Unlike in the case of the $\delta$\,Scuti star HD\,50844 \citep{2009A&A...506...85P}, 
the frequencies do not form a plateau below a given cutoff frequency.

Following the commonly used approach, we considered a frequency as
significant if its amplitude exceeds a S/N ratio of 4
\citep[see][]{1993A&A...271..482B}, while the noise level is calculated as the
average amplitude -- before prewhitening -- in a given interval centered on the
frequency of interest. 
We checked the effect of changing the width of the window in which we
measure the noise levels between the prewhitening steps. Setting the window 
width at $1\,\mathrm{d}^{-1}$ or $3\,\mathrm{d}^{-1}$ resulted in 107 or 149 
frequencies having a S/N level higher than 4 
(confidence level 99.9\% in the case of white Gaussian noise), respectively.
For comparison, a window width over the entire considered frequency range of 
$20\,\mathrm{d}^{-1}$ would result in 394 frequencies with a S/N above 4. 
Changing the significance criterion from 4 to 3.6 or 3 times the residual
amplitude before prewhitening -- corresponding to 95\% or 80\% confidence levels \citep{1997A&A...328..544K}
-- raises the number of significant frequencies from 107--149--394 to
145--186--498 and 297--375--765, respectively. We continued with the 4 S/N
criterion.

Raising the number of frequencies in the model from 107 to 394 lowers the
residual noise levels (calculated in $3\,\mathrm{d}^{-1}$ wide windows centered
around 1.5, 5 and 10 $\mathrm{d}^{-1}$) from 10.4--3.7--1.3 ppm to 1.9--1.8--1.2
ppm (compared to the average signal levels of 92.4--26.9--4.6 in the original
data).  Subtracting a model using 107, 149 or 394 frequencies results in a
variance reduction of 96.27\%, 97.73\%, and 98.93\%, respectively. 

To compare different models, constructed using different sets of parameters, 
calculation of statistics is useful, 
like Akaike's Information Criterion (AIC) and the Bayesian Information Criterion (BIC) 
\citep[see, e.g.,][]{2009A&A...506..111D,2011A&A...533A...4B}. The AIC is defined as
\[\mathrm{AIC} = 2k - 2\ln\mathcal{L}_{\mathrm{max}} \approx n \ln\,(\mathrm{RSS}/n) + 2k + n,\] 
where $\mathcal{L}_{\mathrm{max}}$ is a maximum likelihood estimator (MLE), $n$ is the 
effective number of observations, and $k$ is the number of free parameters in the model. 
Under the assumption of Gaussian white noise, we can insert the MLE of the noise variance, 
$\sigma^2_i = \mathrm{RSS}/n$ with RSS being the residual sum of squares. The BIC is defined as 
\[\mathrm{BIC} = -2\ln\mathcal{L}_{\mathrm{max}}+k \ln n \approx n \ln\,(\mathrm{RSS}/n)+k \ln n.\] Both 
the AIC and the BIC are only relevant in comparisons; the lower their values, the better the model.
Unlike the variance reduction, they discourage the use of 
many free parameters. As shown by \citet{2009A&A...506..111D}, the BIC is more suitable when one wants to stress the 
importance of simpler models over more complicated ones as it increases the penalty 
for introducing new parameters. Although both the AIC or BIC values suggest that raising the
number of free parameters in the model is useful from a viewpoint of these
statistical criteria, it seems meaningless to understand the physics of the
star, as raising to 800 frequencies in the fit only results in an additional
0.26\% of variance reduction.

\subsubsection{Alternative approach}

It is clear that changes in the definitions of the used significance criteria
strongly affect the number of accepted frequencies.  To come to a final list of
frequencies for interpretation, we divided the data set in two equally long time
strings and ran the same iterative prewhitening procedure. The hypothesis
testing for a $p$ value of $p = 0.001$ stopped the process at 736 and 760
frequencies for the first and second half.  We compared the frequency content of
the two parts of the data to each other and to the one of the full data set,
following the matching frequency method described by
\citet{2011A&A...525A..23C}. We set the threshold of having equal frequencies to
the Rayleigh limit of the full data set $1/T=0.00665\,\mathrm{d}^{-1}$ and we
required that amplitudes found in the subsets do not differ more than 50\% from
the amplitude which was found in the complete data set.  Moreover, we also
required that the $n$th prewhitened frequency in the subsets had to be among the
first $n\pm(10\sqrt{n})\,\mathrm{th}$ frequencies from the full data set to
ensure that the order of prewithening in the subsets does not differ too much
from the one in the full set. In this way we found 217 frequencies common for both
halves and the full data set.

A model constructed using these 217 frequencies gives a variance reduction of
96.30\%. To correct for the loss of the lowest frequencies, we completed the
set of common frequencies with those peaks which only appear in the full data
set at S/N$>$4.  This gives an additional 16, 35, or 207 frequencies when a
window width of $1\,\mathrm{d}^{-1}$, $3\,\mathrm{d}^{-1}$, or
$20\,\mathrm{d}^{-1}$ is used to calculate the noise levels. Using these
additional frequencies to construct a model with 233, 252 or 424 frequencies
leads to clearly better fits with a variance reduction of 97.58\%, 98.11\%, or
98.95\%.

Looking at the residual light curves, our approach is more solid than simply
using the classical S/N criterion on the full data set. Thus, we accept the set
of 424 frequencies (listed in Table\,\ref{frequtable}) as the final model of the
light curve. The orbital frequency of the CoRoT satellite was weakly detected, but it did 
not fit the selection criteria. The aliases -- some of which can be easily seen on 
Fig.\,\ref{corotlcurvefourier} -- introduced by the orbit of the CoRoT satellite 
were removed by the iterative prewhitening procedure. 
The noise level in the periodogram calculated from the residuals
around 1.5--5--10$\,\mathrm{d}^{-1}$ is 2.0--1.8--1.1\,ppm, respectively. 
To be conservative and to avoid being completely overwhelmed by the large number of
low-amplitude frequencies due to the large window width for the S/N computation, 
we use the secure set of 252 frequencies 
when we look for characteristic mode frequency spacings in Sect.\,\ref{spacings} 
in order to focus future modelling on those first. We do point out that the additional 
172 lower-amplitude frequencies listed in Table\,\ref{frequtable} also contain 
physical information, e.g., in terms of even more low-amplitude pulsation modes compared 
to those considered primarily or some sort of granulation signal which can be interpreted 
after basic modelling of the 252 dominant ones has been accomplished.

\subsubsection{Notes on the frequency spectrum}

Already upon the first visual inspection of the data we noticed that several
features of the light curve and of the periodogram (see
Fig.\,\ref{corotlcurvefourier}) of HD\,43317 are similar to those of the other
hybrid SPB/$\beta$\,Cep CoRoT star HD\,50230
\citep{2010Natur.464..259D,degroote2011}. HD\,43317 is a single relatively fast
rotator with a projected rotational velocity of over $100\,\mathrm{km\,s}^{-1}$
while this quantity is below $10\,\mathrm{km\,s}^{-1}$ for the primary of the
binary HD\,50230.

As for HD\,50230, we ascribe the lower frequencies (below $2\,\mathrm{d}^{-1}$)
as due to gravity modes, but in the case of HD\,43317 most of the power is
distributed over a smaller number of peaks.  Compared to HD\,50230, we find more
frequencies in the range between $2\,\mathrm{d}^{-1}$ and
$8\,\mathrm{d}^{-1}$. The typical relative amplitudes of these low-order modes
with respect to those of the $g$ modes are higher ($\sim30\%$) than in the case
of HD\,50230 ($\sim7\%$).  Furthermore, these low-order mode frequencies seem to
be members of multiplet structures, distributed over peak groups where the
density of peaks is higher than in neighbouring regions, except for the very
strong and stable isolated peak at $\sim4.33\,\mathrm{d}^{-1}$, which resembles
the one at $\sim4.92\,\mathrm{d}^{-1}$ in the power spectrum of HD\,50230.
Given the similar fundamental parameters and the common features of the two
power spectra, we expect that these isolated peaks have the same physical
origin.

The observed frequencies in a rotating star are related to the pulsational frequency 
in the co-rotating frame by the expression \[f=|f'-mf_{\rm rot}|,\] where $f'$ is the 
frequency in the co-rotating frame and $f_{\rm rot}$ is the rotational frequency of the 
star. For a direct comparison to theoretical predictions, the observed frequency spectrum would need to be 
converted to the co-rotating frame, but this is not possible without knowing the 
azimuthal order of the modes. For fast 
rotators, the difference between the frequency values in the two reference frames can be substantial, and the
observed frequencies will spread over a wider range in the periodogram. This might partly explain
the excess in number and power of the modes we see above $2\,\mathrm{d}^{-1}$ compared to HD\,50230.

\begin{figure}
\resizebox{\hsize}{!}{\includegraphics{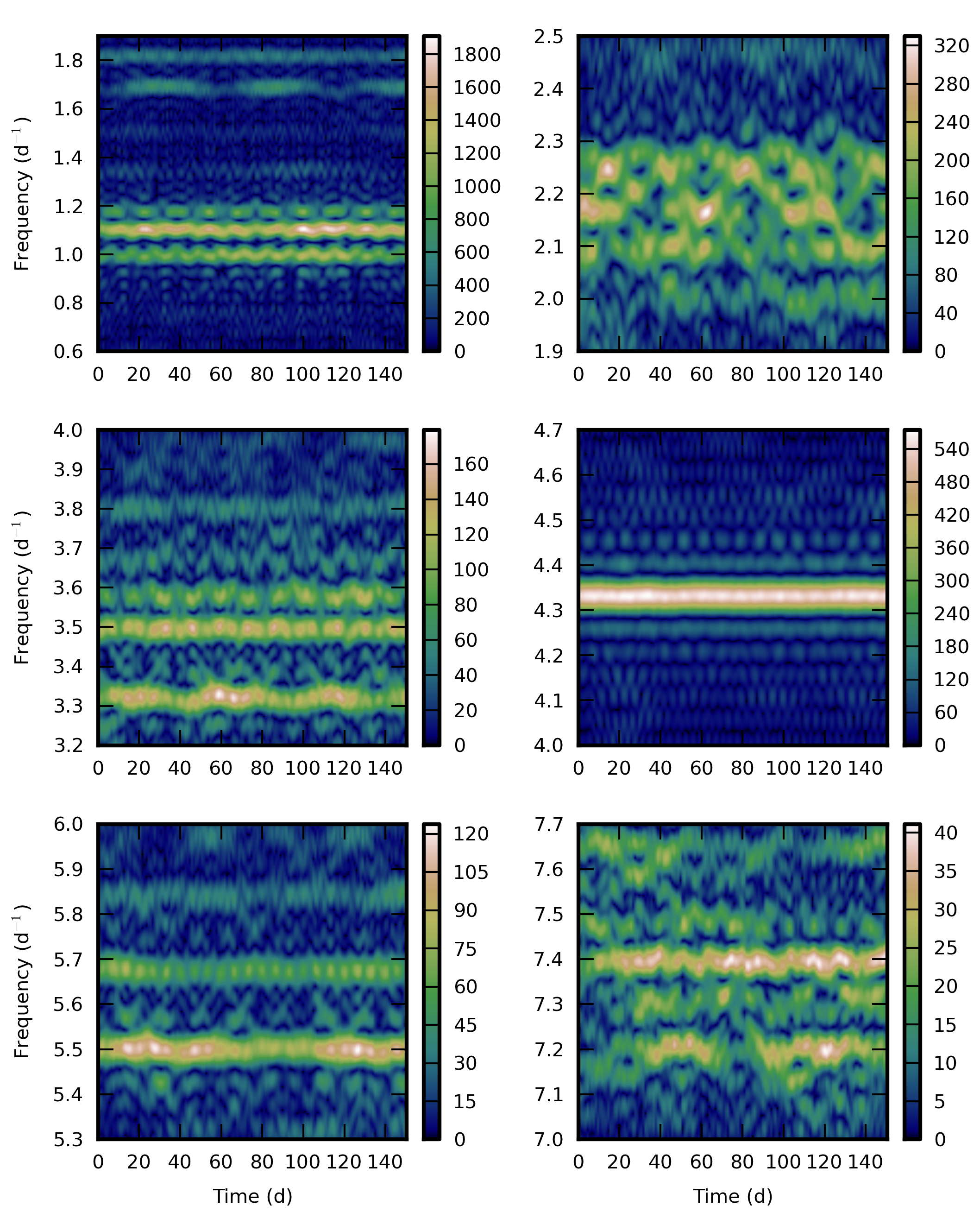}} 
\caption{Short-time Fourier transformations of selected regions of the power
  spectrum (window width of $20\,\mathrm{d}$, colours denote amplitude in
  ppm). All signal outside these regions was removed before the
  calculation. Although closely spaced peaks might introduce complicated beating
  patterns (\textit{upper right}), frequencies appear through the whole light
  curve.}
\label{stft}
\end{figure}

Our frequency analysis method focused on the detection of stable modes with 
long lifetimes as expected for the dominant modes in this type of star 
\citep[e.g.,][]{2010aste.book.....A}. The highest amplitude 
closely spaced peaks in the periodogram of HD\,43317 indeed cause a typical beating
pattern in the short-time Fourier transformation (STFT) (see
Fig.\,\ref{stft}). These frequencies and their amplitude are stable as
prewhitening some of them removes the interference pattern and leaves clear
horizontal structures for the remaining ones in the STFT (even in case of the seemingly complicated
pattern visible on the upper right image on Fig.\,\ref{stft}).

\section{\textsc{Harps} spectroscopy}\label{harpsspectroscopy}
\subsection{Observations and reduction}
In the framework of ground-based preparatory and follow-up observations for the
CoRoT space mission, high-resolution and high S/N spectra were taken
with the \textsc{Harps} \citep{2003Msngr.114...20M}
instrument, in the EGGS mode ($R\approx80\,000$). We refer to Table\,\ref{specsummary} for a summary of the
spectroscopic observations taken simultaneously with the photometric
observations.

\begin{table*}
  \caption{Logbook of the spectroscopic observations of HD\,43317 obtained during December 2009 -- grouped by observing runs.}
\label{specsummary}
\centering
\begin{tabular}{c c c c c c c c}
\hline\hline
Instrument & N & HJD begin & HJD end & $\langle\mathrm{SNR}\rangle$ & SNR-range & $\mathrm{T}_{\mathrm{exp}}$ & R\\
\hline
\textsc{Harps}   & 132 & 2455174.57665 & 2455183.87419 & 248 & [153, 313]& [240, 600]  &$80\,000$\\
\textsc{Harps}   &  59 & 2455191.56933 & 2455195.84863 & 249 & [186, 291]& [250, 500]  &$80\,000$\\
\hline
\end{tabular}
\tablefoot{For each observing run, the instrument, the number of spectra N, the
  HJD of mid-exposure, the average S/N ratio (calculated as the average S/N in the line free regions of [4490\AA -- 4505\AA, 4953\AA -- 4968\AA, 6035\AA -- 6050\AA, 6738\AA -- 6753\AA]), the range of SNR values, the typical exposure times (in seconds), and the resolving power of the spectrograph are given.}
\end{table*}

\begin{figure*}
\resizebox{\hsize}{!}{\includegraphics{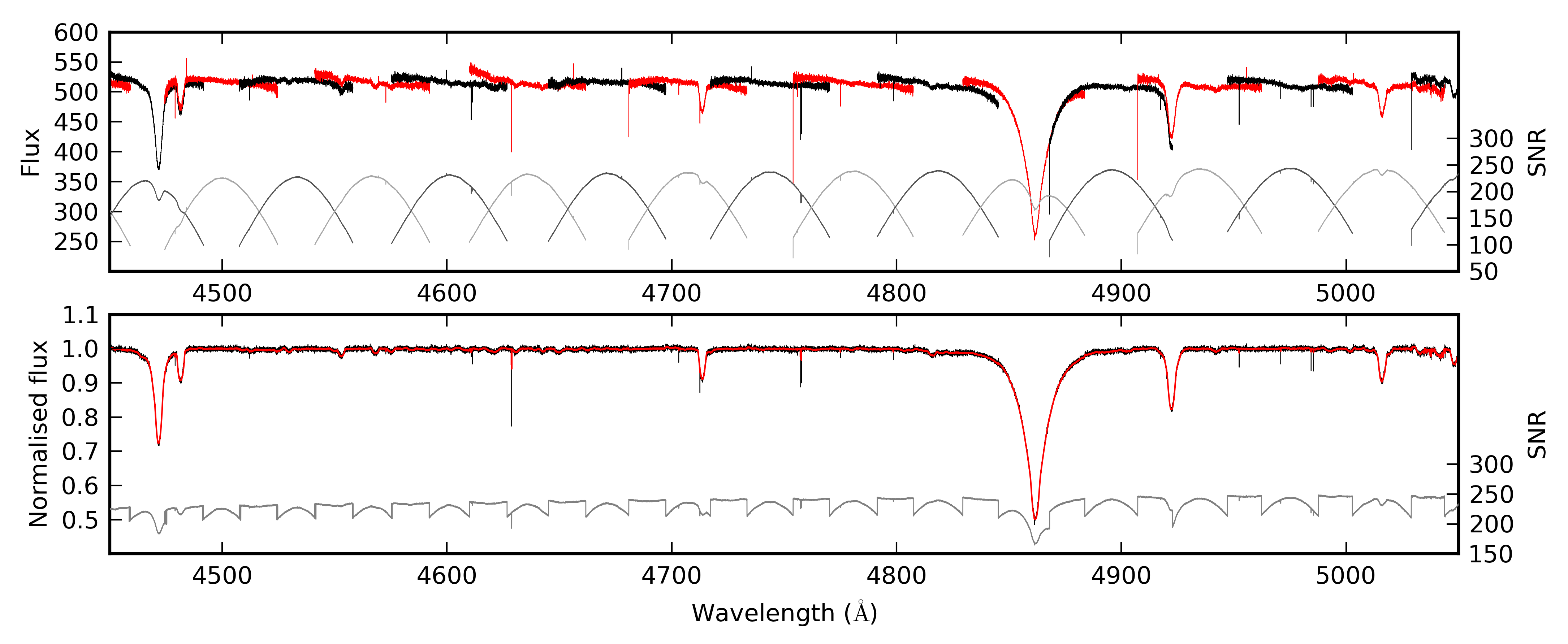}} 
\caption{(\textit{upper panel}) A typical raw \textsc{Harps} spectrum
  (consecutive orders plotted with a different colour for better
  visibility). (\textit{lower panel}) The same spectrum after order merging,
  normalisation, and the removal of cosmics which were not removed by the
  standard pipeline (with the average of all spectra also plotted in
  red). Signal-to-noise levels are also plotted order-by-order and for the
  merged spectrum. Detailed explanation of the process can be found in the
  text.}
\label{normalisation}
\end{figure*}

For a trustworthy abundance analysis (see Sect.\,\ref{abundances}) and for a
study of line profile variations (see Sect.\,\ref{lpv}), the normalisation of
the spectra is at the same time a crucial and far from trivial step to which we
paid special attention.  After order-merging, taking into account the S/N in the
overlapping ranges and correcting for the slightly different flux-levels of the
overlapping orders (see Fig.\,\ref{normalisation}), we performed a
half-automated, two step normalisation of each spectrum. First, after all
spectra were brought to the same flux level by dividing them with their median
intensity, we defined a master function (using the first spectrum) with cubic
splines which were fitted through some tens of points at fixed wavelengths,
where the continuum was known to be free of spectral lines. Then we used this
function to clean all spectra from large-scale artificial features, e.g., waves
in the continuum left there by unsatisfactory flat-fielding. The careful
selection of these nodal points enabled us to construct a function which does
not distort the shape (and especially the wings) of the Balmer lines. To correct
for small scale effects, and a remaining global trend of instrumental origin, we
carried out a second normalisation using more nodal points, which were connected
by linear functions in this phase. The wavelengths of these points were the same
for all spectra, but the normalising function was constructed for each of them
based on the local flux levels, instead of applying a master function
again. Using linear functions at this step enabled us to leave the shape of wide
features -- like the Balmer wings -- untouched (by placing only one nodal point
at the blue and red end of such regions), but still make small-scale corrections
(by using a denser nodal point distribution where it was needed). The described
process is much faster, less subjective, and provides better stability than a
one-by-one manual normalisation. Furthermore, it can easily be applied to SB1
binaries (or SB2 systems after spectral disentangling), by applying the
RV-shifts before normalisation, this way placing the nodal points always at the
same position in the zero velocity frame. We found no sign of binarity for
HD\,43317, so this correction was not necessary.

\subsection{Abundance analysis}\label{abundances}
After normalisation we used different methods to determine the fundamental
parameters and abundances from the average normalised spectrum (see
Fig.\,\ref{normalisation}).  

For a first quick estimate, we carried out a full
grid search with four free parameters ($T_\mathrm{eff}$, $\log g$, $v\,\sin\,i$,
and the metallicity $Z$), using the precalculated BSTAR2006 grid
\citep{2007ApJS..169...83L} and the average \textsc{Harps} spectrum. Chi-square
($\chi^2$) values were calculated using selected regions of \ion{H}{i},
\ion{He}{i}, \ion{Si}{ii}, \ion{Si}{iii}, \ion{C}{ii}, \ion{O}{ii},
\ion{Mg}{ii}, \ion{Fe}{ii}, \ion{S}{ii}, and \ion{Ne}{i} lines. The results
given by the minima of polynomial functions fitted to the $\chi^2$ values, are
listed in Table\,\ref{fundparspectr}. We did not perform extensive error
calculations at this stage, thus the listed errors represent the resolution of
the grid. This method suggests Solar metallicity, but note that metallicity can
only be changed by scaling to the Solar values -- and not element by element --
using this grid.

\subsubsection{Spectral synthesis}\label{lte}
We carried out an LTE-based spectrum analysis, using the range of
$[4060\AA-5780\AA]$ of the average spectrum. We used the LLmodels code 
\citep{2004A&A...428..993S} in its most recent parallel version to compute the
atmosphere models and the SynthV code \citep{1996ASPC..108..198T} to compute
the synthetic spectra. The atmosphere models were computed on a grid in
$T_\mathrm{eff}$, $\log g$, and metallicity with the microturbulence
$\xi_\mathrm{t}$ kept fixed to the standard value of $2\,\mathrm{km\,s}^{-1}$.

In a first step, we computed synthetic spectra on a grid in $T_\mathrm{eff}$,
$\log g$, [M/H], $\xi_\mathrm{t}$, and $v\,\sin\,i$ based on a precalculated
library of atmosphere models. Metallicity in this case means that the abundances
of all metals (chemical elements heavier than He) were scaled by the same factor
compared to the Solar composition.

In the second step, we adjusted the abundances of single chemical elements with
the SynthV code by using the atmosphere models computed for the fixed,
optimum metallicity derived in the first step and by taking the abundance table
corresponding to that metallicity as an initial guess. The abundances were
iterated simultaneously with other parameters like $T_\mathrm{eff}$, $\log g$,
$\xi_\mathrm{t}$, and $v\,\sin\,i$ until the minimum in $\chi^2$ has been found.

In the final step, we did a fine-tuning of the fundamental parameters
($T_\mathrm{eff}$, $\log g$, $\xi_\mathrm{t}$, $v\,\sin\,i$ ) by using the table
with derived abundances and the atmosphere models computed for the metallicity
determined in the first step. A more detailed description as well as tests of
the method can be found in \citet{2011A&A...526A.124L}.

Atmospheric parameters and individual abundances determined in this way are
listed in Tables\,\ref{fundparspectr} and \ref{abundancetable}, along with
results from other methods. Since the abundances of all metals (in column
labelled with LTE/LTE in Table\,\ref{abundancetable}) agree with the derived
metallicity of $[\mathrm{M}/\mathrm{H}]=-0.38\pm0.10$\footnote{relative to the Solar metallicity given by \citet{2005ASPC..336...25A}} within the error bars, we
conclude that there is no need to recompute the atmosphere models assuming
individual chemical composition and, consequently, there is no need to do
re-adjustment of the fundamental parameters.

\subsubsection{Equivalent widths and line ratios}\label{nlte}
To check the dependence of our results on the input physics and different
methods, we also carried out a detailed abundance study based on an extensive
grid of synthetic spectra based on the latest versions of the NLTE
line formation codes DETAIL and SURFACE originally developed by
\citet{1984_butler_phd} and \citet{1981PhDT.......113G}, along with
plane-parallel, fully line-blanketed LTE Kurucz atmospheric models
\citep{1993KurCD..13.....K}. It was shown by \citet{2007A&A...467..295N} that
this hybrid approach is adequate for B stars on the main sequence. For further
details on the grid and the iterative method of analysis along with the error
calculations, we refer to \citet{2006A&A...457..651M}, and discuss only the key
points of our case below.

The projected rotational velocity over $\sim100\,\mathrm{km\,s}^{-1}$ causes
such a large line broadening that is it hard to find lines without significant
blending, which is needed to measure equivalent widths (EWs) 
precisely. The first task was the construction of a proper line list. 
The error values in Table\,\ref{abundancetable} mainly result from the small 
number of lines and the sensitivity of the abundances to changes in the effective temperature. 
We also had to fix the value of
microturbulence at $3 \pm3\,\mathrm{km\,s}^{-1}$, which is a typical value for
main sequence B3 stars \citep[see, e.g.,][]{2010ASPC..425..146N}, because we did
not have enough lines to determine it from requiring independence between the
abundances and the line strengths. Microturbulence is generally found to be
higher only in evolved objects \citep[see, e.g.,][]{2009A&A...499..279C}.

For spectral type B3, we observe only the two ionisation stages \ion{Si}{ii} and
\ion{Si}{iii}, which we used to determine the effective temperature. Here we were 
not only restricted by the need of unblended lines, but also by the fact that some 
lines are not properly modelled \citep{2010A&A...510A..22S}. We fitted
the wings of the $\mathrm{H}\alpha$, $\mathrm{H}\beta$, $\mathrm{H}\gamma$, and
$\mathrm{H}\delta$ lines to fix the $\log g$ value (see
Table\,\ref{fundparspectr}).

To be consistent, we used the same set of lines (8 \ion{He}{i} and 16 metal
lines) as in the abundance analysis to measure the projected rotational
velocity, $v \sin i$, with the Fourier-method described by
\citet{2006A&A...448..351S}. We calculated the $v \sin i$ for all lines
separately, and estimated the error as the standard deviation of these values
(see Table\,\ref{fundparspectr}).

\subsubsection{Discussion of results from different methods}

Compared to the underabundance of metals we found from the spectral synthesis in
Sect.\,\ref{lte}, the second method (see column labelled with LTE/NLTE in
Table\,\ref{abundancetable}) leads to abundances in agreement with the Solar
values within the error bars (except -- although the errors are large -- for carbon). The results for helium are
less clear but the Solar values are still within the error bars for both
analyses.  The low metallicity from the LTE/LTE method would not lead to
excitation \citep[see, e.g., the instability strips calculated for different
metallicities in][]{2007CoAst.151...48M}, while the Solar
abundances from the LTE/NLTE method agree well with similar results found for
other B stars \citep{2009CoAst.158..122M} or the proposed cosmic abundance
standard derived from B stars in the Solar neighbourhood by
\citet{2008ApJ...688L.103P}.  The differences in the resulting abundances and in
the systematic uncertainty of 1000\,K in the effective temperature illustrate
that the LTE treatment for line formation is incorrect in this temperature
range. \citet{2011JPhCS.328a2015P} warns against the blind use of the BSTAR2006 grid, 
as oversimplification in the model atom descriptions might cause problems for some of the complex ions/elements.  
This should not be an issue here, as only a few weak lines of such elements were taken into account.

Given the above, we consider the parameters listed in the last column of Table\,\ref{fundparspectr} (and the corresponding abundances from Table\,\ref{abundancetable}) as the best estimates we can get. As the error in effective temperature is an internal -- and probably unrealistically small -- value, we prefer to adopt a typical error of $\pm1000\,\mathrm{K}$ instead. Using the distance measurement from \citeauthor{2007A&A...474..653V} and the available colour excess values (see Sect.\,\ref{prior}) the estimated $M_\mathrm{V}=-1.40\pm0.51$ is fully consistent -- through the spectral class of B3.5\,V which is based on the derived fundamental parameters -- with our results \citep[based on the tables provided by][]{1981Ap&SS..80..353S}.

\begin{table*}
  \caption{Fundamental stellar parameters derived from spectroscopy using different methods, model atmospheres, and line formation codes.}
\label{fundparspectr}
\centering
\begin{tabular}{l c c c}
\hline\hline
Method	&Quick look grid-search	&Spectral synthesis &Equivalent widths\\
Atmosphere				&NLTE	&LTE		&LTE\\
Line formation			&NLTE	&LTE		&NLTE\\
\hline
$T_\mathrm{eff}\,(\mathrm{K})$ 			&$17\,350 \pm1000$				&$16\,800 \pm100$	&$17\,750 \pm250$\\
$\log g\,(\mathrm{cgs})$					&$4.0 \pm0.25$					&$3.9 \pm0.1$		&$4.1 \pm0.1$\\
$v \sin i\,(\mathrm{km\,s}^{-1})$		&$108 \pm10$						&$106 \pm4$			&$115 \pm9$\\
$\xi_\mathrm{t}\,(\mathrm{km\,s}^{-1})$	&$2$ (preset)					&$3.35 \pm0.80$		&$3 \pm3$ (preset)\\
Spectral type\tablefootmark{a}		&&B4\,V-IV&B3.5\,V\\
\hline
\end{tabular}
\tablefoot{\tablefoottext{a}{Spectral type had been determined based on $T_\mathrm{eff}$ and $\log g$ values by using an interpolation in the tables given by \citet{1982SchmidtKalerBook}.}}
\end{table*}

\begin{table*}
  \caption{Abundances derived from different methods based on different model atmospheres and line formation codes compared to Solar values and a cosmic abundance standard.}
\label{abundancetable}
\centering
\begin{tabular}{l c c c c c c}
\hline\hline
Element	&LTE	/LTE				&LTE	/NLTE (n lines)	&GS98\tablefootmark{a}	&AGS05\tablefootmark{b}	&AGSS09\tablefootmark{c}	&PNB08\tablefootmark{d}\\	
\hline                                                                                                                            						
He		&$11.13\pm0.20$		&$10.78\pm0.27\,(8)$	&$10.93\pm0.004$			&$10.93\pm0.01$			&$10.93\pm0.01$				&$10.98\pm0.02$\\			
C		&$8.14 \pm0.20$		&$8.00 \pm0.28\,(2)$	&$8.52 \pm0.06$			&$8.39 \pm0.05$			&$8.43 \pm0.05$				&$8.32 \pm0.03$\\			
N		&					&$7.66 \pm0.28\,(2)$	&$7.92 \pm0.06$			&$7.78 \pm0.06$			&$7.83 \pm0.05$				&$7.76 \pm0.05$\\			
O		&$8.11 \pm0.20$		&$8.77 \pm0.24\,(2)$	&$8.83 \pm0.06$			&$8.66 \pm0.05$			&$8.69 \pm0.05$				&$8.76 \pm0.03$\\			
Ne		&					&$7.99 \pm0.12\,(5)$	&$8.08 \pm0.06$			&$7.84 \pm0.06$			&$7.93 \pm0.10$				&$8.08 \pm0.03$\\			
Mg		&$7.18 \pm0.20$		&$7.23 \pm0.39\,(1)$	&$7.58 \pm0.05$			&$7.53 \pm0.09$			&$7.60 \pm0.04$				&$7.56 \pm0.05$\\			
Si		&$6.86 \pm0.20$		&$7.45 \pm0.26\,(4)$	&$7.55 \pm0.05$			&$7.51 \pm0.04$			&$7.51 \pm0.03$				&$7.50 \pm0.02$\\			
S		&$6.74 \pm0.20$		&					&$7.33 \pm0.11$			&$7.14 \pm0.05$			&$7.12 \pm0.03$				&\\							
Fe		&$6.80 \pm0.20$		&					&$7.50 \pm0.05$			&$7.45 \pm0.05$			&$7.50 \pm0.04$				&$7.44 \pm0.04$\\		
\hline
\end{tabular}
\tablefoot{All abundances are given in units of $\log(\mathrm{El/H})+12$ / atoms per $10^6$ nuclei. \tablefoottext{a}{\citet{1998SSRv...85..161G}; }\tablefoottext{b}{\citet{2005ASPC..336...25A}; }\tablefoottext{c}{\citet{2009ARA&A..47..481A}; }\tablefoottext{d}{\citet{2008ApJ...688L.103P}.}}
\end{table*}

\subsection{Study of line profile variations}\label{lpv}

Pulsations result in radial velocity changes and variations in spectral line
profiles. Depending on the degree $\ell$ and azimuthal order $m$, modes leave
characteristic patterns in the spectroscopic
data, which can be used for mode identification if we are able to decipher
them. In order to achieve this, we use the pixel-by-pixel method to detect
the pulsation frequencies from the variation in the line profiles and use the
Fourier-parameter fit (FPF) method \citep{2006A&A...455..227Z} to identify the
modes, as well as the line profiles' moments \citep{2003A&A...398..687B}. We
have investigated our spectroscopic data using both methods, but the broad
lines imply the pixel-by-pixel method to be best suited.

As discussed by \citet[][Chapter 4 and Chapter 6]{2010aste.book.....A}, it is best to perform 
spectroscopic mode identification from one unblended deep line, whose width is dominated
by thermal broadening and not affected by pressure broadening. For the case of HD\,43317,
the \ion{He}{i} line at 6678\AA\ and the \ion{Mg}{ii} line at 4481\AA\ fulfill these
requirements best. We thus treat them first. Further on, we discuss the variability 
behaviour of other spectral lines.

\subsubsection{The line moments}

\begin{table}
\caption{Frequencies derived from the Fourier analysis of the line moments.}
\label{moments}
\centering
\begin{tabular}{c c c}
\hline\hline
						&\multicolumn{2}{c}{Frequency $(\mathrm{d}^{-1})$}\\
						&\ion{He}{i} at 6678\AA 	&\ion{Mg}{ii} at 4481\AA\\
\hline
$\langle v \rangle$		&$2.23$					&\\
						&						&$1.35$\\
						&$1.10$					&$1.10$\\
						&						&$1.69$\\
\hline
$\langle v^2 \rangle$	&$2.23$					&\\
						&$4.33$					&$4.33$\\
\hline
$\langle v^3 \rangle$	&$2.23$					&\\
						&$1.10$					&$1.10$\\
						&$1.69$					&$1.69$\\

\hline
\end{tabular}
\tablefoot{The errors of the frequencies are $1/T_{\mathrm{spectr.}}\approx0.05\,\mathrm{d}^{-1}$.}
\end{table}

\begin{figure}
\resizebox{\hsize}{!}{\includegraphics{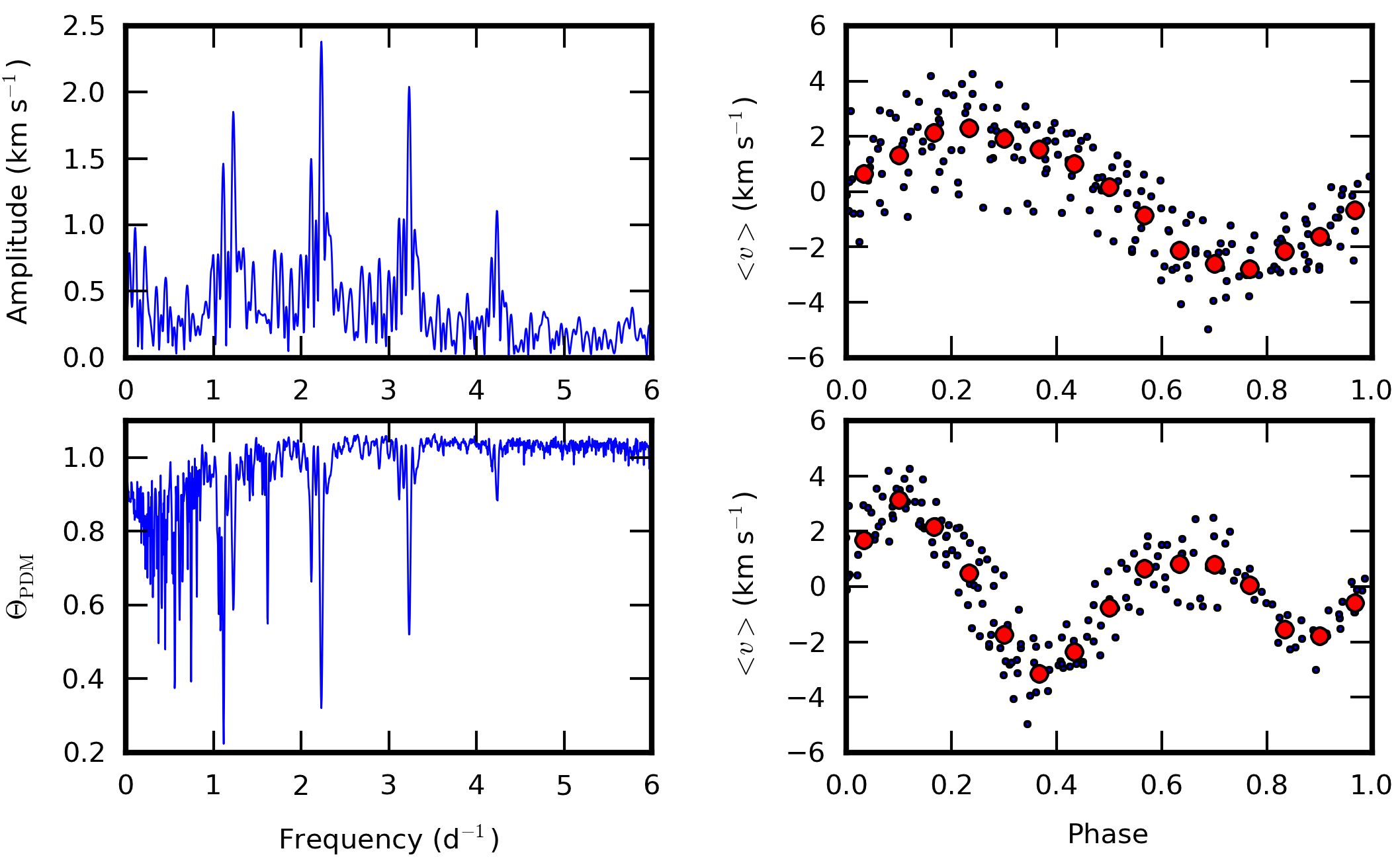}} 
\caption{(\textit{upper left}) Scargle periodogram of $\langle v \rangle$ of the
  \ion{He}{i} line at 6678\AA\, with the highest peak at
  $\sim2.23\,\mathrm{d}^{-1}$. (\textit{upper right}) Phase-plot of $\langle v
  \rangle$ according to the frequency $f_\mathrm{m}=2.228748\,\mathrm{d}^{-1}$.
  The small symbols are observations, while the larger red circles are
  phase-binned averages. (\textit{lower left})
  $\Theta_{\mathrm{PDM}}$-statistics of $\langle v \rangle$, showing the deepest
  minimum at $f_\mathrm{m}/2$. (\textit{lower right}) Phase-plot of $\langle v
  \rangle$ according to the frequency $f_\mathrm{m}/2$.}
\label{momentplots}
\end{figure}

We calculated the first three moments ($\langle v \rangle$, $\langle v^2
\rangle$, and $\langle v^3 \rangle$) for the selected helium and magnesium lines and
computed their discrete Fourier-transform function (DFT) using the FAMIAS
software package \citep{2008CoAst.155...17Z}.  The results are listed in
Table\,\ref{moments}.  All frequencies in Table\,\ref{moments} match significant
peaks found in the CoRoT dataset (see Table\,\ref{frequtable}). Given that the
frequency precision from the spectroscopy is an order of magnitude lower
(Rayleigh frequency of $1/T_{\mathrm{spectr.}}\approx0.05\,\mathrm{d}^{-1}$),
the match is not always unambiguous.

The frequency content for the two spectral lines is not completely
overlapping. There is a frequency which can be only found in the moments of the
\ion{He}{i} line at 6678\AA\ ($\sim2.23\,\mathrm{d}^{-1}$), and there is a
frequency which can be found in the \ion{Mg}{ii} line at 4481\AA\ alone
($\sim1.35\,\mathrm{d}^{-1}$).

The highest peak in the DFT of the first moment of the He line is
$\sim2.23\,\mathrm{d}^{-1}$ and its phase-plot shows a sine pattern with
moderate scatter. However, the phase dispersion minimisation \citep[PDM,
see][]{1978ApJ...224..953S} technique suggests that the true frequency is half
the value found from the DFT, which clearly leads to a lower scatter in the
phase diagram (Fig.\,\ref{momentplots}). The EW of the \ion{He}{i}
line at 6678\AA\ shows the same behaviour as the first moment (with a small
phase shift between the two). The Mg line does not show such double-wave nature.

\subsubsection{Variability in other selected lines}

Following \citet{2010aste.book.....A} we investigated if we could increase the
significance of the line profile variability in a global sense, by applying the 
Least Square Deconvolution technique \citep[LSD,][]{1997MNRAS.291..658D,1998ApJ...495..440K}.
In order to do so, we started from the set of unblended lines selected in Sect.\,\ref{abundances}.
Such methods only deliver useful and better results than individual lines if their variability 
has the same amplitude, phase, and intrinsic broadening for all the profiles considered in the merging.
Given their strong Stark broadening, H lines are excluded for such tests while He lines must be
treated separately from the metal lines. Small blends will lead to uncertainties, especially when the number of
lines used for the calculation is low, and must be avoided.

We calculated two sets of LSD profiles, one using only helium lines, and another
one based on metal lines, and compared their behaviour to the one of the \ion{He}{i}
line at 6678\AA\ and of the \ion{Mg}{ii} line at 4481\AA.
In case of the \ion{He}{i} lines, the LSD profiles are slightly
asymmetric as the individual lines have small blends in their wings.
Even though the line depths are the same, and the signal-to-noise of the LSD profiles is far higher than the one of 
the single \ion{He}{i} line at 6678\AA, its variability amplitude is only half that of the 
individual line. In the case of the metal lines, the LSD profiles are shallower, and 
even though their signal-to-noise levels are higher, their variability signal is again only
half that of the individual \ion{Mg}{ii} line at 4481\AA.

The derived frequencies for both sets are identical to those found in the
individual lines and we do not find any additional ones. This shows that, also 
globally in the spectrum, the \ion{He}{i} lines behave differently than the metal lines,
except for the \ion{S}{ii} lines (e.g., at 5640\AA) and the 
\ion{Ne}{i} lines (e.g., at 6096\AA, 6143\AA, 6267\AA, 6402\AA, and 6507\AA) which show similar
behaviour to the \ion{He}{i} lines. The rest of the metal lines with clear line-profile variability,
e.g., \ion{C}{ii} at 4267\AA, \ion{Si}{ii} at 5056\AA, 6347\AA, 6371\AA, \ion{Si}{iii} at 4553\AA, and 4568\AA,
behave in a similar manner to the selected \ion{Mg}{ii} line, but add to the complexity in the LSD due to small blends
and various levels of thermal broadening.

We conclude that we could not improve the results of the frequency detection and mode
identification from the LSD technique, which is not really surprising in the case of hot stars given
their limited number of spectral lines.

\subsubsection{Pixel-by-pixel variations}\label{pixelbypixelstudy}

\begin{table}
\caption{Frequencies derived from the pixel-by-pixel analysis of the line profiles.}
\label{pixelbypixel}
\centering
\begin{tabular}{c c c c c}
\hline\hline
\multicolumn{2}{c}{Frequency $(\mathrm{d}^{-1})$}	&$l$		&$m$		&$i\,(^{\circ})$\\
\ion{He}{i} at 6678\AA 	&\ion{Mg}{ii} at 4481\AA	&		&		&\\
\hline
$2.23$					&						&		&		&\\
$1.36$					&$1.35$					&$3$		&$3$		&$29$\\
$4.45$					&						&		&		&\\
$3.58$					&$3.58$					&$8,7$	&$0$		&$34$\\
$4.33$					&$4.33$					&$2\,(\geq2)$	&$2\,(\geq0)$	&$23\,(20-40)$\\
$6.68$					&						&		&		&\\
$1.11$?					&						&		&		&\\
						&$2.76$?					&		&		&\\
\hline
\end{tabular}
\tablefoot{The errors of the frequencies are $1/T_{\mathrm{spectr.}}\approx0.05\,\mathrm{d}^{-1}$. Values with a question mark are non-secure frequencies. For the discussion of mode identification see Sect.\,\ref{spectrinterpretation}.}
\end{table}

\begin{figure}
\resizebox{\hsize}{!}{\includegraphics{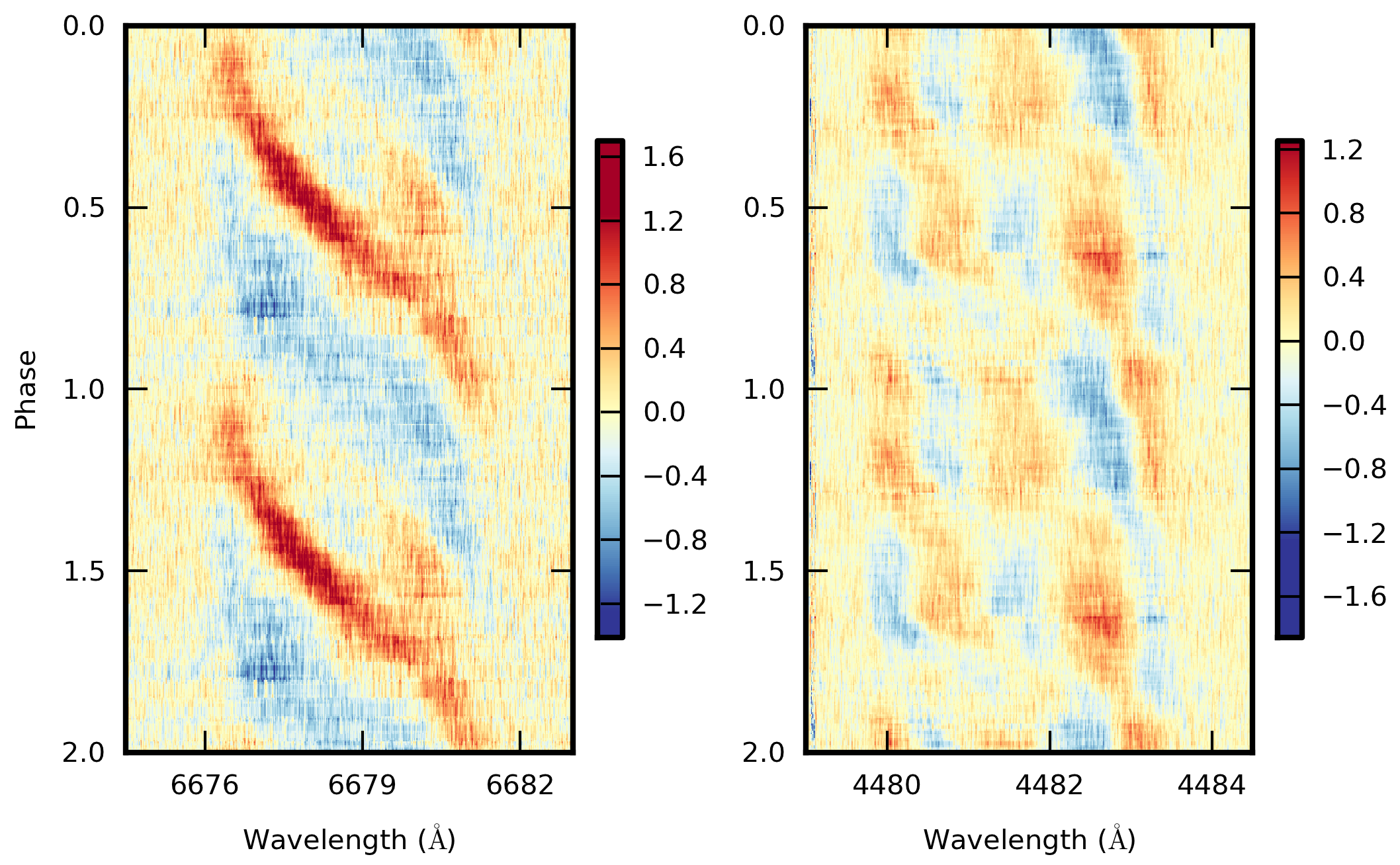}} 
\caption{A colour scale representation of the residual spectral lines with
  respect to the average line for the selected He and Mg line of HD\,43317.  The
  phase corresponds to the dominant period found through the pixel-by-pixel
  method, colours denote deviations from the average profile in percentage. For
  better visibility of the patterns, two cycles are shown, and we have applied
  an S/N-weighted sliding boxcar smoothing with a bin width of 0.05 and a step
  size of 0.005 cycles. (\textit{left}) Phase-plot for the \ion{He}{i} 
  line at 6678\AA\ using the dominant frequency of
  $2.228748\,\mathrm{d}^{-1}$(\textit{right}) Phase-plot for the \ion{Mg}{ii}
  line at 4481\AA\ using the dominant frequency of $1.354276\,\mathrm{d}^{-1}$.}
\label{lpvfigure}
\end{figure}

The frequencies found with the pixel-by-pixel method are listed in
Table\,\ref{pixelbypixel}.  Just as in case of the line moments, all frequencies
can be matched to significant peaks found in the CoRoT photometry. For both
spectral lines, the dominant frequency matches the one found in the line
moments. Phase-plots of the line profiles according to these frequencies are
shown in Fig.\,\ref{lpvfigure}.

All frequencies derived from the \ion{Mg}{ii} line at 4481\AA\ are also found in
the pixel-by-pixel variations of the \ion{He}{i} line at 6678\AA.  The additional
frequencies in the He line are $\sim2.23\,\mathrm{d}^{-1}$,
$\sim4.45\,\mathrm{d}^{-1}$, and $\sim6.68\,\mathrm{d}^{-1}$.  The frequency
$\sim1.11\,\mathrm{d}^{-1}$ is also present in the residuals, but it does not
meet our significance criterion of having a S/N level above 4. However, all the
frequencies which can only be found in the variations of the He line are even
harmonics of $\sim1.11\,\mathrm{d}^{-1}$.

\subsubsection{Interpretation and discussion}\label{spectrinterpretation}

We interpret the spectroscopic variability as due to rotation combined with
pulsations.  The harmonics detected across the He line, as well as in its EW and
first moment, point towards a spotted or at least inhomogeneous surface
configuration. The fact that such a harmonic structure is not detected in the Mg
line points out that we are dealing with temperature and/or chemical 
variations, to which the He (and Ne)
lines are more sensitive than the Mg line.

\begin{figure}
\resizebox{\hsize}{!}{\includegraphics{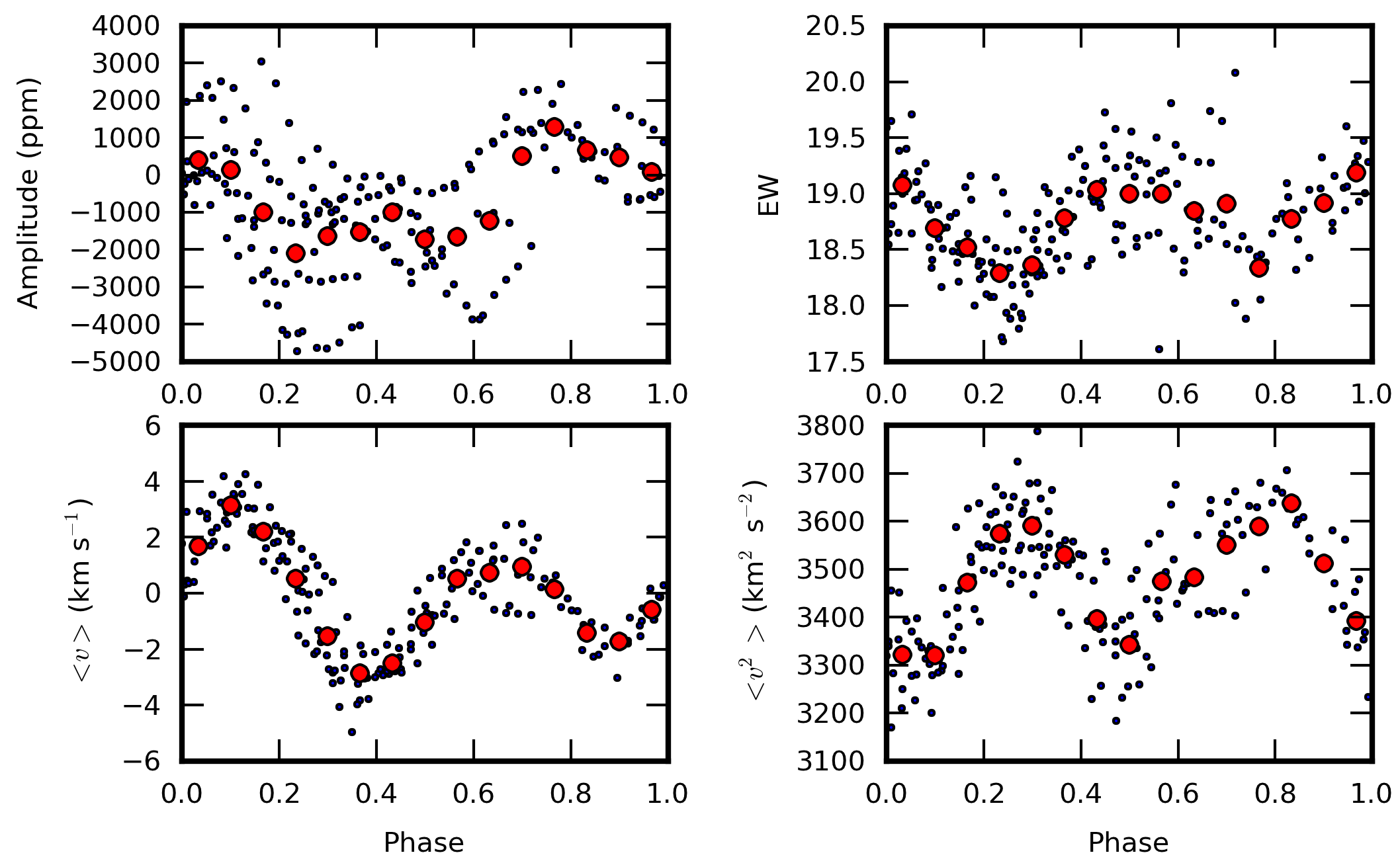}} 
\caption{Phase-plots using the rotation frequency $f_{\mathrm{rot}} = 1.114997\,\mathrm{d}^{-1}$. 
  (\textit{upper left}) The subset of CoRoT data averaged over the \textsc{Harps} integration times 
  when simultaneous spectroscopy is available. The cycle-to-cycle variations are due to 
  the dominant pulsation modes. (\textit{upper
   right}) the EW; (\textit{lower panels}) $\langle v \rangle$ and $\langle v^2
  \rangle$ of the \ion{He}{i} line at 6678\AA. The small symbols
  are observations, while the larger red circles are phase-binned averages.}
\label{moremomentplots}
\end{figure}

The spectroscopic variability leads to the conclusion that $\sim
1.11\,\mathrm{d}^{-1}$ is the rotation frequency of the star. Indeed, the
variations in the moments of the \ion{He}{i} line at 6678\AA\ are best explained
in terms of rotational modulation and the resulting double bump in the
phase-plots (Fig.\,\ref{moremomentplots}) was already detected and interpreted
as such before in other B stars as well.  The moments of HD\,43317 have the same
characteristics as those of the B6 star HD\,105382
\citep[see][]{2001A&A...380..177B,2004A&A...413..273B} while they are atypical
for pulsations \citep[see][]{1992A&A...266..294A,2003A&A...407.1029C}.

\begin{figure}
\resizebox{\hsize}{!}{\includegraphics{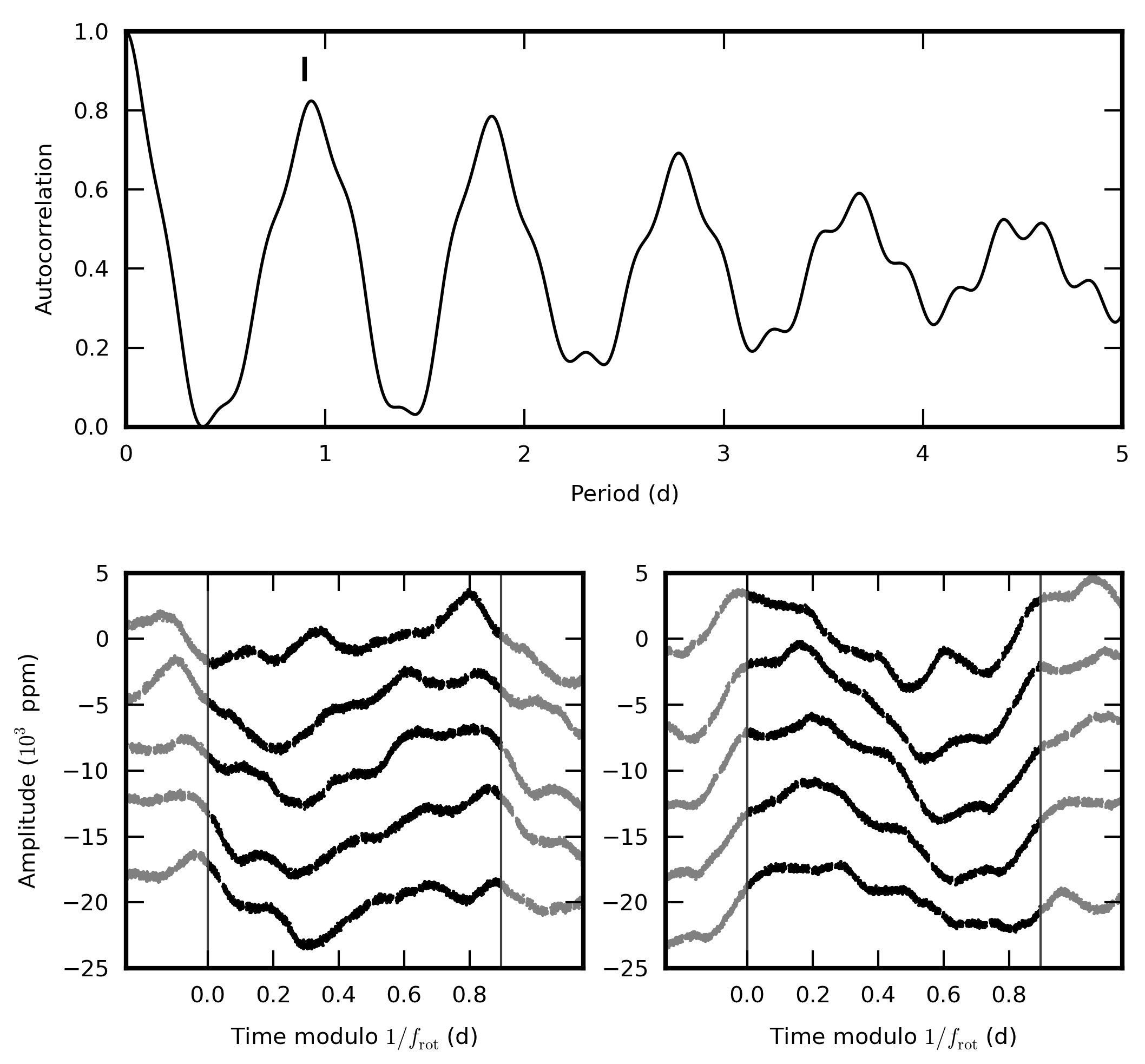}} 
\caption{(\textit{upper panel}) Autocorrelation function of the CoRoT light
  curve. The period corresponding to the proposed $f_\mathrm{rot}$ rotational
  frequency is marked with a thick vertical line. The same scaling is used as
  for Fig.\,\ref{autocorrelation}. (\textit{lower panels}) Sample consecutive
  cuts of the CoRoT light curve according to a period of $1/f_\mathrm{rot}$
  showing strong similarities. For better visibility the previous and following
  cuts are also plotted in grey, while the consecutive orders are shifted downwards 
  by 5000 ppm.}
\label{lightcurveautocorr}
\end{figure}

Furthermore, the autocorrelation function of the CoRoT light curve gives the
strongest peak around a period of $\sim 0.9$\,d (Fig.\,\ref{lightcurveautocorr})
and not half this period. The rotation frequency is not the highest peak in
the Scargle periodogram of the CoRoT light curve ($1.100615\,\mathrm{d}^{-1}$),
but does occur among the largest-amplitude signals at $f_{\rm
  rot}=1.114997\,\mathrm{d}^{-1}$. This is indeed a perfect match with
$f_\mathrm{m}/2$ where $f_\mathrm{m}=2.228748$\,d$^{-1}$ is adopted from the
Fourier-analysis of $\langle v \rangle$ of the \ion{He}{i} line at 6678\AA.  The
frequency $f_{\rm rot}$ is found as a significant peak very close to the dominant
pulsation frequency even when the light curve is cut into two halves.

Further evidence supporting $f_{\rm rot}$ to correspond with the rotation
frequency comes from the fact that a series of harmonics for $f_{\mathrm{rot}}$
was detected up to $7f_{\mathrm{rot}}$ in the CoRoT light curve (these are
marked in Table\,\ref{frequtable}) while no harmonics were found for the
numerous other frequencies.  The amplitude of the harmonics of $f_{\rm rot}$
follows a decreasing trend, with the even harmonics having higher amplitude than
the odd ones above $2f_{\mathrm{rot}}$. The systematic effect that the observed
frequencies differ slightly from the exact harmonics was also found in the
case of the B0.5\,IV star HD\,51756 \citep{2011A&A...528A.123P}, where it was
explained in terms of differential rotation.  The simultaneous occurrence of
rotational modulation and pulsations was also found in the case of the CoRoT
B2.5\,V target HD\,48977 \citep{Thoul2011}, but the frequency resolution is
less for that star.

From models along the evolutionary tracks with input physics in \citet{2011A&A...527A.112B}, 
we consider a typical value of the radius of $R = 3.8R_{\odot}$ for $T_\mathrm{eff}=17\,000\,\mathrm{K}$, 
$\log g = 4.0\,\mathrm{(cgs)}$, and a metallicity of $Z=0.014$. This results in 
$v_{\mathrm{eq}}\approx214\,\mathrm{km\,s}^{-1}$. This, combined with the
measured projected rotational velocity $v \sin i \approx110\,\mathrm{km\,s}^{-1}$, gives an 
inclination angle $i\approx30^{\circ}$. A corresponding typical mass in the models is 
$\mathcal{M}=5.4\mathcal{M}_{\odot}$ and leads to a critical velocity of 
$v_{\mathrm{cr}}\approx425\,\mathrm{km\,s}^{-1}$ \citep[in the definition of][]{1995ApJ...440..308C}. 
These radius and mass values, and the deduced critical velocity, are averages for the considered 
model grid and are in agreement with the ones in the calibration tables of \citet{1981Ap&SS..80..353S} 
for the spectral type B3.5\,V which we derived in Sect.\,\ref{abundances}. 
Thus, the star is rotating at $\sim50\%$ of its critical velocity. Since we do not detect any signs 
of emission in the spectrum, as expected for a star rotating at the break-up limit and thus losing material, 
it is unlikely that $f_\mathrm{m}$ is the rotation frequency.

We restricted the estimated inclination value for the mode identification ($i\in
[20^{\circ}$,$40^{\circ}]$) using the FPF method in FAMIAS, which takes into
account rotations effects due to the Coriolis force in the modelling of the
pulsations  \citep[for the
practical details, we refer to][]{2006A&A...455..227Z,2008CoAst.155...17Z}. We
applied it to the three common modes which were found in the pixel-by-pixel
analysis of both data sets in Sect.\,\ref{pixelbypixelstudy}. The results are
shown in Table\,\ref{pixelbypixel}. For the frequency at
$\sim1.36\,\mathrm{d}^{-1}$, we performed the analysis on both selected spectral
lines, with the same result.  For the two other frequencies, we used only the Mg
line, as the influence from the harmonics of the rotational frequency was too
high in the He line, causing the residual signal to be much weaker in that
case. Based on both the $\chi^2$ values and the visual check of the amplitude
and phase fits, we are confident in the mode identification of $\ell=m=3$ for the
peak at $\sim1.36\,\mathrm{d}^{-1}$, while for the two other peaks our level of
confidence is lower, especially in case of the frequency at
$\sim4.33\,\mathrm{d}^{-1}$. Although the best fit is achieved with $\ell=m=2$,
the quality of the fit itself is not very convincing, especially given the large
spread in $\ell$ and $m$ within a small range in $\chi^2$.

We have to note that all the available mode identification methods, including
those implemented in FAMIAS, are based on a perturbative treatment of the
rotation of the star. This is doubtful in the case of gravito-inertial modes
\citep{2010A&A...518A..30B}, i.e., modes for which $f<2f_{\rm rot}$. All the
frequencies in Table\,\ref{frequtable} with value below $2.23\,\mathrm{d}^{-1}$
are in that regime, in particular the dominant one at
$\sim1.36\,\mathrm{d}^{-1}$. In that sense, also the identification of that
dominant mode must be treated as preliminary.

Given that the Coriolis force changes the propagation cavities for
gravito-inertial modes compared to those of g modes with $f>2f_{\rm rot}$, it
might be that the dominant mode's excitation, and the one of all the modes in
that regime, were affected by the rotation of the star. This could be an
explanation why the relative strength of the g modes in this star with respect
to the p modes, is different than in the case of the slow rotator HD\,50230.

\section{Search for characteristic spacings}\label{spacings}

Although fitting individual frequency values is generally thought to deliver
good constraints on the models of massive stars \citep[see,
e.g.,][]{2004MNRAS.355..352A}, it relies on the assumption that there is no
offset between observed and computed oscillation frequencies in the case of
stars with a radiative envelope.  In order to be less dependent of this
assumption, it is worthwhile to search for useful seismic diagnostics to model,
such as frequency and/or period spacings, as well as rotational splittings.

\begin{figure}
\resizebox{\hsize}{!}{\includegraphics{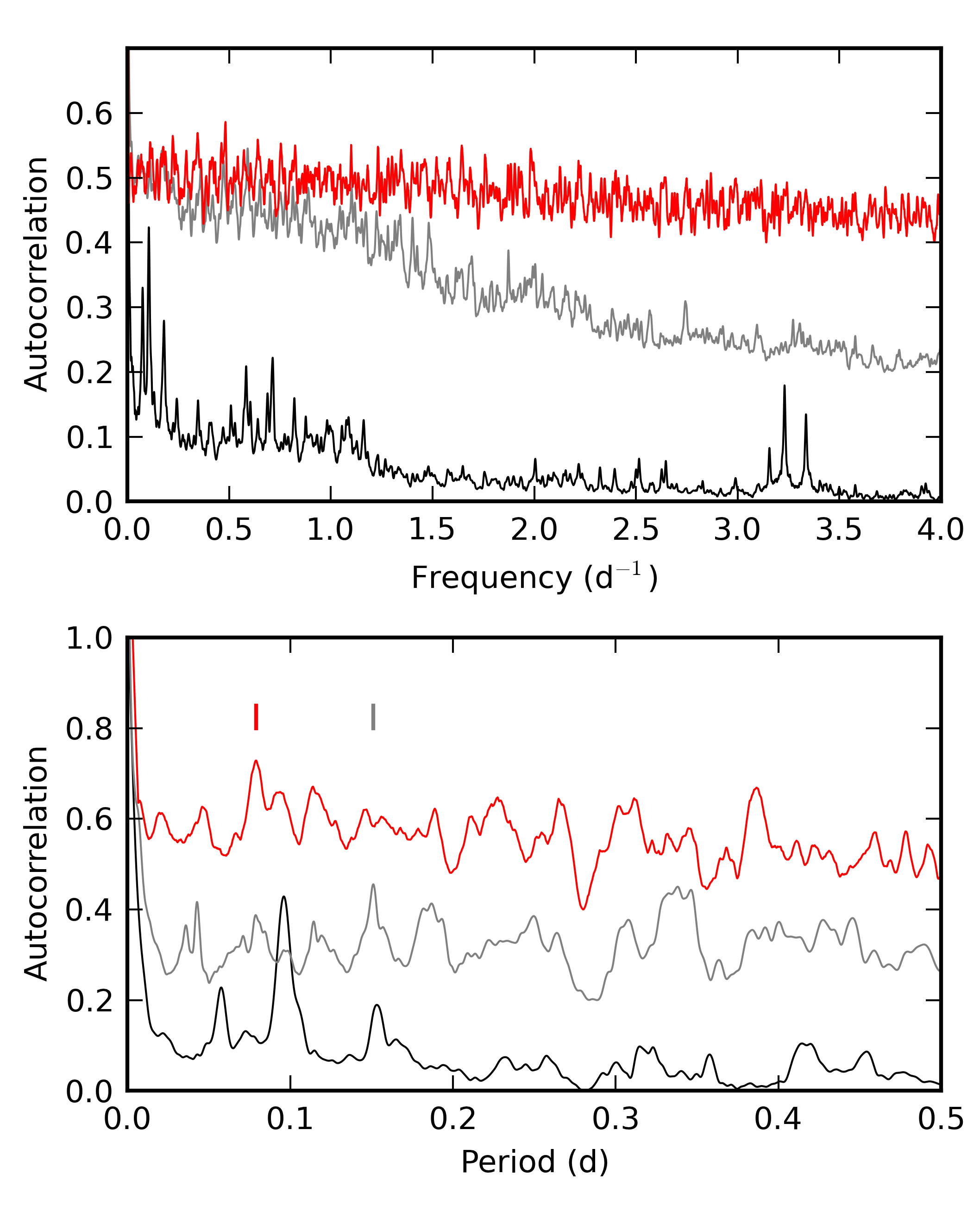}} 
\caption{(\textit{upper panel}) Autocorrelation function of the full power
  spectrum in frequency space. The autocorrelation of the original power
  spectrum is plotted in black, the one of the power spectrum after prewhitening
  the 6 strongest peaks is plotted in grey, and then one of the 252 significant
  frequencies given an artificial power while all others given zero power in
  red.  (\textit{lower panel}) Autocorrelation functions of the power spectrum
  between $0.5\,\mathrm{d}^{-1}$ and $2\,\mathrm{d}^{-1}$, converted into period
  space (colours have the same meaning).  Candidate spacings are marked with a
  thick vertical line using a colour according to the corresponding function.}
\label{autocorrelation}
\end{figure}

The autocorrelation function of the power spectrum of the CoRoT light curve is
shown in Fig.\,\ref{autocorrelation}.  The few strong peaks each correspond with
the frequency difference between two of the dominant modes.  The autocorrelation
function computed after the six strongest modes have been removed, no longer
shows prominent peaks and has similar behaviour as the one we obtained from
the spectrum containing the 252 significant frequencies when all these modes
were given the same artificial power and all other frequencies more than half
the Rayleigh frequency away from the 252 under consideration were placed at
value zero.

Although the autocorrelation functions (see Fig.\,\ref{autocorrelation}) do not
result in apparent spacing values, we find not only triplets, but also series of
more components.  We considered all candidate spacings in the data, starting
from pairs to longer series, using the set of 252 accepted frequencies, for a
tolerance set to the Rayleigh frequency. After having found candidates, we 
confronted them with the autocorrelation functions.
 
It is hard to exclude rotational splitting as interpretation of a spacing
without knowing the $\ell$ and $m$ values.  In fact, rotational splittings of
HD\,43317 can easily be asymmetric due to the relatively high rotational
frequency, and we may have missed rotational multiplets due to this reason with
our criteria.

\begin{figure}
\resizebox{\hsize}{!}{\includegraphics{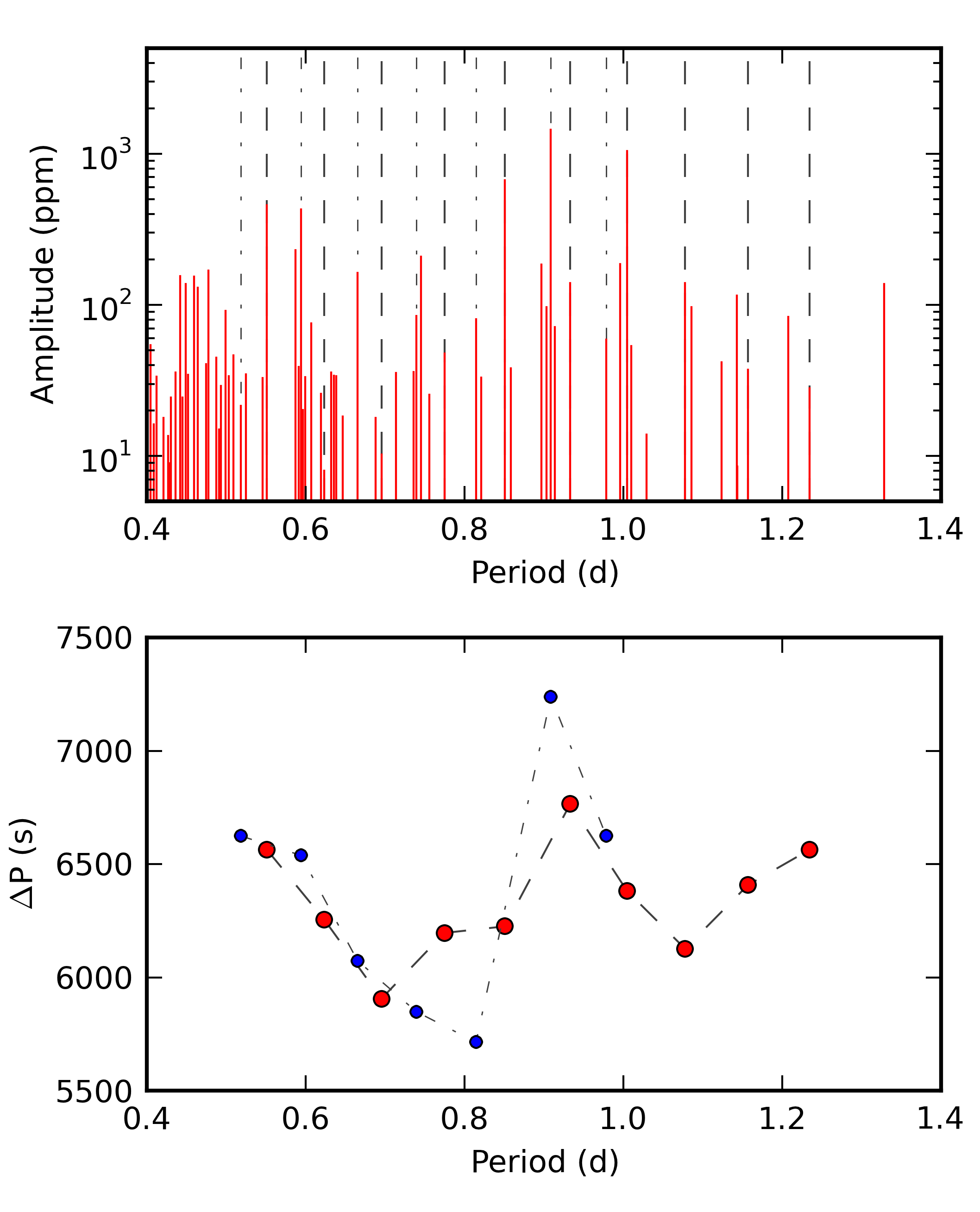}} 
\caption{(\textit{Upper panel}) The period spectrum of the CoRoT light curve of
  HD\,43317 (red solid lines), with the locations of the members of the two
  longest series with almost constant period spacing in grey dashed, and
  dash-dotted lines. (\textit{Lower panel}) The deviations from the average
  period spacings are indicated through line connections between the members of
  the two series.}
\label{periodspacing}
\end{figure}

To find appropriate candidates for frequency and period chains, we looked for
series having a high average amplitude, a relatively small scatter (standard
deviation) of amplitudes, and containing many peaks.  The longest series we
found in this way is one with ten consecutive peaks (see
Fig.\,\ref{periodspacing}), having an average period spacing of 6339\,s
(0.07337\,d). This value is physically meaningful for B3\,V stars (e.g., Aerts \&
Dupret 2011) and is of the same order as the spacing value found for the slow
rotator HD\,50230 \citep[9418\,s,][]{2010Natur.464..259D}. As in that paper, we
also find small deviations from the exact spacing value, with an average
absolute deviation of 197\,s (0.00228\,d). In addition to this longest series of
consecutive peaks, we find another one with seven components, whose peaks do not
overlap with the previous series and having an average period spacing of 6380\,s
(0.07385\,d), with a deviation of 430\,s (0.00498\,d). This series is also
indicated on Fig.\,\ref{periodspacing}. 

The dominant spectroscopic mode is not a member of the series found. This is as
expected, given that it has an $\ell=3$ while we expect the two detected long
series of peaks from the photometry to correspond with $\ell=1$ or 2 modes
having the same azimuthal order within the series. In fact, given that 
the found period 
spacing of the two series is about the same suggests that we are dealing with
one $\ell$-value and different $m$-values. It is not a critical issue
that we are unable to convert these frequencies to a co-rotating frame
without the secure knowledge of their azimuthal orders, as 
from a modelling point of view the frequency spacings of the series are of main importance, 
and the transformation between the two reference frames does not affect these
values.

Additional shorter sequences of frequencies also occur, but they contain less
consecutive peaks and are thus more prone to randomness. They could be suitable
for interpretation once the main sequences of frequencies can be well identified
and fitted by model frequencies.


\section{Conclusions}

Our observational seismic study of the fast rotator HD\,43317 by means of
CoRoT space photometry and high-precision \textsc{Harps} spectroscopy led to its
classification as a new bright hybrid $\beta\,$Cep/SPB star. The star rotates at
about half its critical velocity and shows clear evidence of rotational
modulation in addition to its pulsational variability in both the photometry and
the spectroscopy. The list of B-type stars with a detected mixture of rotational
and pulsational variability with similar frequencies and amplitudes is growing
and occurs across all effective temperatures of this spectral type \citep[e.g.,][]
{2005A&A...432..955U, 2011A&A...536A..82D, Thoul2011}. This is
not surprising given the co-existence of chemically peculiar Bp and SPB stars in
the same part of the HR diagram \citep[e.g.,][]{2007A&A...466..269B}.

With the CoRoT target HD\,43317, we deliver the second pulsating B star for
which almost constant period spacings of its high-order g modes have been
detected.  As shown by \citet{2010Natur.464..259D}, this offers the potential to
tune the details of the conditions, in particular the level and nature of
mixing, near the stellar core of such stars.  In the case of HD\,50230, this was
done ignoring the effects of the slow rotation of the star on the behaviour of
its g modes, a valid assumption for that star given its equatorial rotation
velocity below 10\,km\,s$^{-1}$. HD\,43317 has a factor ten higher rotational
frequency and we are thus in a more complex situation.  Interpretation of the
detected period spacings near 6300\,s and their small deviation in terms of the
input physics of stellar models, as it was done in the case of the slow rotator
HD\,50230 \citep[9418\,s,][]{2010Natur.464..259D}, first requires secure
identification of the wavenumbers $(\ell,m)$ of the modes corresponding to the
frequency peaks indicated in Fig.\,\ref{periodspacing}. In particular, we must
unravel if all the peaks in one series correspond with one value of the
azimuthal order $m$, and, if so, with which one. Given that we were unable to
detect those frequencies in \textsc{Harps} spectroscopy, such an identification
must come from predictions of oscillation modes from theoretical stellar
models. In computing such pulsation predictions, it must be taken into account
that many of the detected gravity modes of HD\,43317 fall in the regime of the
gravito-inertial modes, whose propagation cavities are modified by the Coriolis
force. This is in particular the case for all the 17 frequencies occurring in
the two series with the constant period spacing we found. The description of
such modes requires full computations with respect to the rotation of the star,
rather than approximations based on a perturbative approach.


\begin{acknowledgements}
  The research leading to these results has received funding from the European
  Research Council under the European Community's Seventh Framework Programme
  (FP7/2007--2013)/ERC grant agreement n$^\circ$227224 (PROSPERITY), as well as
  from the Belgian Science Policy Office (Belspo, C90309: CoRoT Data
  Exploitation). EP and MR acknowledge financial support from the PRIN-INAF 2010
  \textit{Asteroseismology: looking inside the stars with space- and ground-based 
  observations}. TM acknowledges financial support from Belspo for contract 
  PRODEX-GAIA DPAC. MH acknowledges support by the Austrian Fonds zur F\"orderung 
  der wissenschaftlichen Forschung (FWF, project P\,22691-N16).
\end{acknowledgements}

\bibliographystyle{aa}
\bibliography{HD43317}


\Online
\begin{appendix}
\section{Tables}\label{tables}

\begin{longtable}{c c c c c c c c c l}
  \caption{\label{frequtable} Fourier parameters (frequencies ($f_j$),
    amplitudes ($A_j$), and phases ($\theta_j$)) of peaks which are common in
    both halves of the data (marked with a 1 in the note column), as well as
    additional ones which have a S/N ratio above 4 when computed over the full
    periodogram. The displayed S/N values are calculated in a window of
    $1\,\mathrm{d}^{-1}$ or $3\,\mathrm{d}^{-1}$ centred on the given frequency,
    or from the full
    periodogram (from $0\,\mathrm{d}^{-1}$ to $20\,\mathrm{d}^{-1}$).}\\
  \hline\hline
  $f\,(\mathrm{d}^{-1})$ & $\epsilon_f\,(\mathrm{d}^{-1})$ & $A\,(\mathrm{ppm})$ & $\epsilon_A\,(\mathrm{ppm})$ & $\theta\,(2\pi/\mathrm{rad})$ & $\epsilon_{\theta}\,(2\pi/\mathrm{rad})$ & \multicolumn{3}{c}{S/N ($1\,\mathrm{d}^{-1}$, $3\,\mathrm{d}^{-1}$, full)} & notes\\
  \hline
\endfirsthead
\caption{continued.}\\
\hline\hline
$f\,(\mathrm{d}^{-1})$ & $\epsilon_f\,(\mathrm{d}^{-1})$ & $A\,(\mathrm{ppm})$ & $\epsilon_A\,(\mathrm{ppm})$ & $\theta\,(2\pi/\mathrm{rad})$ & $\epsilon_{\theta}\,(2\pi/\mathrm{rad})$ & \multicolumn{3}{c}{S/N ($1\,\mathrm{d}^{-1}$, $3\,\mathrm{d}^{-1}$, full)} & notes\\
\hline
\endhead
\hline
\endfoot
  0.009115 &   0.000052 &    47.6 &     0.7 &  -0.4770 &   0.0142 &   3.3 &   5.1 &  10.8 &      \\ 
  0.013771 &   0.000029 &   164.8 &     1.2 &  -0.2569 &   0.0080 &   5.7 &  10.2 &  22.9 &      \\
  0.017623 &   0.000066 &    39.7 &     0.6 &   0.0120 &   0.0179 &   2.9 &   4.3 &   8.7 &      \\
  0.024001 &   0.000037 &    95.4 &     0.9 &   0.3338 &   0.0100 &   4.5 &   7.8 &  16.8 &      \\
  0.030803 &   0.000047 &    64.3 &     0.8 &   0.2103 &   0.0128 &   3.5 &   5.8 &  12.4 &     1\\
  0.035753 &   0.000104 &    16.4 &     0.5 &  -0.3486 &   0.0284 &   2.8 &   3.2 &   5.7 &      \\
  0.042276 &   0.000067 &    34.0 &     0.6 &   0.2764 &   0.0183 &   2.8 &   4.2 &   8.4 &      \\
  0.048332 &   0.000085 &    23.4 &     0.5 &  -0.4108 &   0.0232 &   2.7 &   3.5 &   6.7 &      \\
  0.059474 &   0.000052 &    52.6 &     0.7 &  -0.0151 &   0.0141 &   3.4 &   5.3 &  11.2 &      \\
  0.071204 &   0.000126 &    12.5 &     0.4 &  -0.1331 &   0.0343 &   2.8 &   3.0 &   4.8 &      \\
  0.076214 &   0.000096 &    18.4 &     0.5 &   0.4196 &   0.0261 &   2.9 &   3.3 &   6.0 &     1\\
  0.085479 &   0.000128 &    12.4 &     0.4 &   0.2358 &   0.0349 &   2.9 &   3.0 &   4.8 &      \\
  0.093037 &   0.000083 &    24.1 &     0.6 &   0.1388 &   0.0227 &   2.7 &   3.6 &   6.9 &      \\
  0.100324 &   0.000060 &    38.2 &     0.7 &   0.2186 &   0.0164 &   3.0 &   4.7 &   9.6 &      \\
  0.109413 &   0.000080 &    28.5 &     0.6 &  -0.0664 &   0.0219 &   2.5 &   3.7 &   7.1 &      \\
  0.118490 &   0.000060 &    43.3 &     0.7 &  -0.3243 &   0.0164 &   3.0 &   4.7 &   9.7 &     1\\
  0.125854 &   0.000078 &    28.0 &     0.6 &   0.3793 &   0.0212 &   2.5 &   3.8 &   7.4 &      \\
  0.134565 &   0.000083 &    26.2 &     0.6 &  -0.4583 &   0.0225 &   2.6 &   3.6 &   6.9 &      \\
  0.142858 &   0.000156 &     9.5 &     0.4 &   0.1679 &   0.0425 &   2.8 &   2.7 &   4.2 &      \\
  0.151574 &   0.000099 &    18.5 &     0.5 &  -0.0448 &   0.0270 &   2.8 &   3.3 &   5.9 &      \\
  0.171872 &   0.000051 &    54.7 &     0.8 &   0.3362 &   0.0140 &   3.3 &   5.4 &  11.4 &     1\\
  0.183878 &   0.000096 &    18.9 &     0.5 &  -0.2657 &   0.0263 &   2.7 &   3.3 &   6.0 &      \\
  0.190282 &   0.000140 &    11.0 &     0.4 &  -0.3417 &   0.0381 &   2.8 &   2.8 &   4.5 &      \\
  0.199595 &   0.000065 &    37.4 &     0.6 &   0.2156 &   0.0179 &   2.9 &   4.3 &   8.7 &      \\
  0.205877 &   0.000044 &    67.7 &     0.8 &   0.1483 &   0.0121 &   3.7 &   6.3 &  13.5 &     1\\
  0.216482 &   0.000103 &    17.0 &     0.5 &  -0.1845 &   0.0281 &   2.8 &   3.3 &   5.7 &      \\
  0.223234 &   0.000023 &   227.3 &     1.4 &  -0.1935 &   0.0063 &   7.7 &  13.2 &  30.1 &     1\\
  0.229468 &   0.000077 &    25.5 &     0.6 &  -0.4640 &   0.0209 &   2.6 &   3.8 &   7.4 &      \\
  0.234981 &   0.000158 &     9.1 &     0.4 &  -0.0538 &   0.0430 &   2.8 &   2.7 &   4.1 &      \\
  0.254685 &   0.000121 &    13.4 &     0.4 &  -0.2498 &   0.0329 &   2.8 &   3.0 &   5.0 &      \\
  0.269158 &   0.000093 &    21.1 &     0.5 &  -0.1011 &   0.0254 &   2.6 &   3.4 &   6.2 &     1\\
  0.280472 &   0.000128 &    12.7 &     0.4 &  -0.4197 &   0.0348 &   2.8 &   3.0 &   4.8 &      \\
  0.287154 &   0.000088 &    22.9 &     0.5 &  -0.3232 &   0.0240 &   2.7 &   3.5 &   6.5 &      \\
  0.297694 &   0.000088 &    24.8 &     0.5 &  -0.4673 &   0.0241 &   2.6 &   3.4 &   6.5 &      \\
  0.312549 &   0.000067 &    38.3 &     0.7 &   0.0775 &   0.0183 &   2.7 &   4.2 &   8.6 &      \\
  0.324384 &   0.000119 &    13.5 &     0.4 &   0.4872 &   0.0325 &   2.7 &   3.0 &   5.1 &      \\
  0.347868 &   0.000132 &    12.8 &     0.4 &   0.3285 &   0.0359 &   2.8 &   2.9 &   4.7 &     1\\
  0.366114 &   0.000055 &    48.6 &     0.7 &  -0.2562 &   0.0151 &   3.2 &   5.0 &  10.5 &     1\\
  0.372783 &   0.000125 &    12.6 &     0.4 &  -0.2347 &   0.0340 &   3.0 &   3.1 &   5.0 &      \\
  0.390784 &   0.000040 &    83.6 &     0.9 &  -0.0452 &   0.0110 &   4.1 &   7.0 &  15.1 &     1\\
  0.404773 &   0.000093 &    21.0 &     0.5 &   0.0330 &   0.0254 &   2.7 &   3.4 &   6.2 &      \\
  0.412766 &   0.000080 &    21.7 &     0.6 &  -0.4832 &   0.0218 &   2.6 &   3.6 &   7.1 &      \\
  0.419957 &   0.000105 &    16.4 &     0.5 &  -0.1856 &   0.0286 &   2.8 &   3.2 &   5.6 &     1\\
  0.443743 &   0.000094 &    19.4 &     0.5 &   0.3643 &   0.0258 &   2.7 &   3.3 &   6.1 &      \\
  0.457698 &   0.000106 &    17.1 &     0.5 &  -0.1683 &   0.0290 &   2.9 &   3.2 &   5.6 &      \\
  0.469130 &   0.000058 &    45.2 &     0.7 &   0.3791 &   0.0159 &   3.0 &   4.8 &   9.9 &     1\\
  0.478903 &   0.000154 &     9.7 &     0.4 &   0.3772 &   0.0421 &   2.8 &   2.7 &   4.2 &      \\
  0.487055 &   0.000163 &     8.8 &     0.4 &   0.4596 &   0.0444 &   2.8 &   2.7 &   4.1 &      \\
  0.493330 &   0.000119 &    13.7 &     0.5 &   0.3145 &   0.0323 &   2.7 &   3.0 &   5.1 &      \\
  0.504681 &   0.000083 &    26.7 &     0.6 &  -0.1294 &   0.0226 &   2.6 &   3.6 &   6.9 &     1\\
  0.509579 &   0.000142 &    11.3 &     0.4 &   0.2877 &   0.0387 &   2.8 &   2.8 &   4.5 &      \\
  0.516619 &   0.000153 &     9.9 &     0.4 &  -0.0216 &   0.0418 &   2.8 &   2.7 &   4.2 &      \\
  0.529073 &   0.000037 &    97.2 &     1.0 &  -0.3303 &   0.0100 &   4.5 &   7.8 &  17.1 &     1\\
  0.534974 &   0.000070 &    31.1 &     0.6 &  -0.4934 &   0.0191 &   2.7 &   4.1 &   8.1 &      \\
  0.544156 &   0.000105 &    17.2 &     0.5 &   0.4816 &   0.0285 &   2.7 &   3.2 &   5.6 &      \\
  0.560246 &   0.000127 &    12.3 &     0.4 &   0.2047 &   0.0346 &   2.9 &   3.0 &   4.8 &      \\
  0.566155 &   0.000103 &    17.3 &     0.5 &   0.4292 &   0.0281 &   2.7 &   3.3 &   5.7 &      \\
  0.576665 &   0.000090 &    21.5 &     0.5 &  -0.3717 &   0.0245 &   2.6 &   3.4 &   6.4 &      \\
  0.583709 &   0.000051 &    55.0 &     0.8 &  -0.2874 &   0.0140 &   3.5 &   5.4 &  11.5 &     1\\
  0.596798 &   0.000095 &    20.2 &     0.5 &   0.3315 &   0.0259 &   2.6 &   3.3 &   6.1 &      \\
  0.602744 &   0.000049 &    56.6 &     0.8 &   0.3487 &   0.0134 &   3.7 &   5.6 &  11.9 &     1\\
  0.611127 &   0.000141 &    10.6 &     0.4 &   0.1055 &   0.0384 &   3.0 &   2.9 &   4.5 &      \\
  0.631848 &   0.000098 &    19.1 &     0.5 &   0.1048 &   0.0267 &   2.6 &   3.3 &   5.9 &      \\
  0.646877 &   0.000082 &    26.3 &     0.6 &  -0.2169 &   0.0224 &   2.6 &   3.6 &   6.9 &     1\\
  0.655913 &   0.000119 &    13.7 &     0.4 &  -0.0626 &   0.0325 &   2.8 &   3.0 &   5.1 &     1\\
  0.664701 &   0.000136 &    11.8 &     0.4 &  -0.0877 &   0.0371 &   2.9 &   2.9 &   4.6 &      \\
  0.674818 &   0.000127 &    11.7 &     0.4 &  -0.1300 &   0.0346 &   2.8 &   3.0 &   4.8 &      \\
  0.680434 &   0.000066 &    38.3 &     0.6 &   0.0452 &   0.0180 &   2.8 &   4.3 &   8.6 &      \\
  0.685828 &   0.000066 &    38.9 &     0.6 &   0.1027 &   0.0180 &   3.0 &   4.3 &   8.7 &      \\
  0.691662 &   0.000046 &    62.9 &     0.8 &   0.0407 &   0.0126 &   4.0 &   6.0 &  12.8 &     1\\
  0.703770 &   0.000119 &    14.0 &     0.4 &  -0.2372 &   0.0326 &   2.8 &   3.0 &   5.1 &      \\
  0.714386 &   0.000097 &    18.7 &     0.5 &   0.2333 &   0.0265 &   2.7 &   3.3 &   6.0 &      \\
  0.722836 &   0.000148 &    10.0 &     0.4 &  -0.3729 &   0.0405 &   2.9 &   2.8 &   4.3 &      \\
  0.736937 &   0.000150 &    10.4 &     0.4 &  -0.2879 &   0.0409 &   2.8 &   2.7 &   4.3 &      \\
  0.746246 &   0.000109 &    15.0 &     0.5 &   0.1262 &   0.0299 &   2.7 &   3.2 &   5.4 &      \\
  0.752946 &   0.000029 &   137.4 &     1.1 &   0.2209 &   0.0079 &   5.9 &   9.9 &  22.1 &     1\\
  0.763685 &   0.000083 &    25.2 &     0.6 &   0.2844 &   0.0227 &   2.7 &   3.6 &   6.9 &      \\
  0.772414 &   0.000095 &    19.0 &     0.5 &  -0.1277 &   0.0260 &   2.9 &   3.4 &   6.1 &      \\
  0.784377 &   0.000141 &    11.0 &     0.4 &   0.0008 &   0.0384 &   2.8 &   2.8 &   4.5 &      \\
  0.797230 &   0.000098 &    20.2 &     0.5 &  -0.3280 &   0.0266 &   2.7 &   3.3 &   6.0 &      \\
  0.809990 &   0.000070 &    28.1 &     0.6 &   0.0829 &   0.0192 &   2.9 &   4.1 &   8.1 &     1\\
  0.820705 &   0.000124 &    13.1 &     0.4 &  -0.1828 &   0.0339 &   2.7 &   3.0 &   4.9 &      \\
  0.828119 &   0.000040 &    83.1 &     0.9 &   0.2946 &   0.0108 &   4.6 &   7.1 &  15.3 &     1\\
  0.834297 &   0.000084 &    24.2 &     0.6 &   0.1241 &   0.0230 &   2.7 &   3.5 &   6.8 &      \\
  0.848231 &   0.000159 &     9.1 &     0.4 &  -0.1634 &   0.0433 &   2.8 &   2.7 &   4.1 &      \\
  0.857879 &   0.000102 &    17.3 &     0.5 &   0.1599 &   0.0279 &   2.7 &   3.3 &   5.8 &      \\
  0.864453 &   0.000066 &    37.4 &     0.7 &   0.0941 &   0.0179 &   3.2 &   4.3 &   8.8 &      \\
  0.874755 &   0.000167 &     8.5 &     0.4 &   0.4974 &   0.0456 &   3.0 &   2.8 &   4.0 &     1\\
  0.874931 &   0.000032 &   114.9 &     1.0 &   0.1521 &   0.0088 &   5.2 &   8.9 &  19.5 &     1\\
  0.882486 &   0.000088 &    23.5 &     0.5 &   0.4802 &   0.0239 &   2.7 &   3.5 &   6.5 &      \\
  0.890244 &   0.000061 &    41.7 &     0.7 &   0.4405 &   0.0165 &   3.3 &   4.6 &   9.5 &      \\
  0.896303 &   0.000126 &    13.1 &     0.4 &  -0.2383 &   0.0343 &   2.9 &   3.0 &   4.9 &      \\
  0.901933 &   0.000075 &    29.0 &     0.6 &   0.4692 &   0.0205 &   2.7 &   3.9 &   7.6 &      \\
  0.920948 &   0.000036 &    96.4 &     1.0 &  -0.4238 &   0.0100 &   4.9 &   7.8 &  17.1 &      \\
  0.928011 &   0.000029 &   139.6 &     1.2 &  -0.3092 &   0.0080 &   5.7 &  10.0 &  22.4 &     1\\
  0.933250 &   0.000092 &    21.1 &     0.5 &  -0.2815 &   0.0250 &   2.6 &   3.4 &   6.3 &      \\
  0.938338 &   0.000159 &     9.1 &     0.4 &  -0.1061 &   0.0434 &   2.8 &   2.7 &   4.1 &      \\
  0.945177 &   0.000080 &    25.3 &     0.6 &  -0.3440 &   0.0219 &   2.6 &   3.6 &   7.1 &      \\
  0.953175 &   0.000111 &    15.5 &     0.5 &   0.3027 &   0.0302 &   2.7 &   3.1 &   5.4 &      \\
  0.971876 &   0.000119 &    13.8 &     0.4 &  -0.0812 &   0.0325 &   2.7 &   3.0 &   5.1 &     1\\
  0.982627 &   0.000083 &    24.0 &     0.6 &  -0.3811 &   0.0227 &   2.6 &   3.5 &   6.9 &      \\
  0.990351 &   0.000051 &    53.4 &     0.8 &   0.1158 &   0.0139 &   4.0 &   5.4 &  11.4 &      \\
  0.995475 &   0.000008 &  1043.5 &     2.4 &   0.3215 &   0.0023 &  20.7 &  42.1 &  98.5 &     1\\
  0.997878 &   0.000096 &    18.8 &     0.5 &   0.2382 &   0.0261 &   2.5 &   3.3 &   6.0 &      \\
  1.003970 &   0.000028 &   185.7 &     1.3 &   0.3484 &   0.0076 &   6.3 &  10.7 &  24.3 &     1\\
  1.008972 &   0.000128 &    12.4 &     0.4 &  -0.4166 &   0.0349 &   3.0 &   3.0 &   4.8 &      \\
  1.016716 &   0.000161 &     8.7 &     0.4 &  -0.3018 &   0.0440 &   2.9 &   2.7 &   4.1 &      \\
  1.022344 &   0.000050 &    59.0 &     0.8 &  -0.2498 &   0.0135 &   3.9 &   5.6 &  11.9 &     1\\
  1.030297 &   0.000123 &    12.9 &     0.4 &  -0.1751 &   0.0337 &   2.9 &   3.0 &   4.9 &      \\
  1.044684 &   0.000099 &    21.8 &     0.5 &  -0.0124 &   0.0271 &   2.7 &   3.2 &   5.8 &      \\
  1.053373 &   0.000101 &    18.3 &     0.5 &   0.4177 &   0.0276 &   2.8 &   3.3 &   5.8 &      \\
  1.060585 &   0.000087 &    23.0 &     0.5 &   0.4089 &   0.0238 &   2.6 &   3.5 &   6.6 &      \\
  1.071783 &   0.000030 &   139.2 &     1.1 &   0.3317 &   0.0081 &   5.9 &   9.8 &  21.9 &     1\\
  1.082798 &   0.000127 &    12.5 &     0.4 &   0.4202 &   0.0347 &   3.0 &   3.0 &   4.8 &      \\
  1.094722 &   0.000044 &    71.2 &     0.8 &   0.2313 &   0.0119 &   4.3 &   6.3 &  13.7 &     1\\
  1.100615 &   0.000007 &  1443.4 &     2.9 &  -0.1757 &   0.0020 &  21.3 &  51.3 & 119.8 &     1\\
  1.107152 &   0.000036 &    96.6 &     0.9 &  -0.2764 &   0.0099 &   5.2 &   7.9 &  17.1 &      \\
  1.114997 &   0.000028 &   185.3 &     1.4 &   0.0490 &   0.0076 &   6.2 &  10.9 &  24.9 &     1, $f_\mathrm{rot}$\\
  1.127400 &   0.000112 &    16.2 &     0.5 &  -0.4907 &   0.0306 &   2.8 &   3.1 &   5.3 &      \\
  1.143003 &   0.000119 &    14.0 &     0.4 &   0.3409 &   0.0325 &   2.8 &   3.0 &   5.1 &      \\
  1.149946 &   0.000093 &    20.5 &     0.5 &  -0.2295 &   0.0253 &   2.6 &   3.4 &   6.2 &      \\
  1.165051 &   0.000062 &    38.1 &     0.7 &   0.3327 &   0.0168 &   3.1 &   4.5 &   9.3 &     1\\
  1.175372 &   0.000011 &   668.1 &     2.1 &  -0.1391 &   0.0031 &  16.8 &  30.3 &  70.8 &     1\\
  1.182695 &   0.000083 &    25.5 &     0.6 &  -0.0670 &   0.0227 &   2.6 &   3.5 &   6.9 &      \\
  1.200957 &   0.000118 &    14.8 &     0.5 &   0.2687 &   0.0322 &   2.7 &   3.0 &   5.1 &      \\
  1.218201 &   0.000068 &    33.1 &     0.6 &   0.0560 &   0.0184 &   3.0 &   4.2 &   8.5 &     1\\
  1.228100 &   0.000040 &    80.5 &     0.9 &   0.0697 &   0.0109 &   4.7 &   7.1 &  15.2 &     1\\
  1.247736 &   0.000140 &    10.6 &     0.4 &  -0.0414 &   0.0381 &   2.8 &   2.8 &   4.5 &      \\
  1.267245 &   0.000165 &     8.9 &     0.4 &   0.3357 &   0.0451 &   3.0 &   2.8 &   4.1 &      \\
  1.283539 &   0.000116 &    14.3 &     0.5 &   0.1708 &   0.0318 &   2.9 &   3.0 &   5.1 &      \\
  1.291179 &   0.000055 &    47.6 &     0.7 &   0.0937 &   0.0151 &   3.6 &   5.0 &  10.5 &     1\\
  1.296678 &   0.000100 &    17.3 &     0.5 &  -0.1632 &   0.0274 &   2.8 &   3.3 &   5.8 &      \\
  1.304188 &   0.000154 &     9.9 &     0.4 &  -0.2397 &   0.0419 &   2.8 &   2.8 &   4.2 &      \\
  1.314294 &   0.000156 &    10.7 &     0.4 &  -0.1301 &   0.0425 &   2.7 &   2.7 &   4.2 &      \\
  1.323855 &   0.000082 &    25.4 &     0.6 &  -0.3843 &   0.0223 &   2.8 &   3.6 &   7.0 &     1\\
  1.342371 &   0.000025 &   208.3 &     1.4 &   0.1328 &   0.0069 &   7.1 &  12.1 &  27.6 &     1\\
  1.352903 &   0.000039 &    84.6 &     0.9 &  -0.3013 &   0.0107 &   5.0 &   7.2 &  15.4 &      \\
  1.359125 &   0.000065 &    36.0 &     0.7 &  -0.2061 &   0.0178 &   3.1 &   4.3 &   8.8 &      \\
  1.366117 &   0.000088 &    22.3 &     0.5 &   0.1675 &   0.0241 &   2.8 &   3.5 &   6.5 &      \\
  1.372954 &   0.000120 &    13.6 &     0.4 &   0.4794 &   0.0329 &   3.0 &   3.0 &   5.0 &      \\
  1.380267 &   0.000105 &    17.4 &     0.5 &  -0.3620 &   0.0287 &   2.9 &   3.2 &   5.7 &      \\
  1.392000 &   0.000163 &     8.9 &     0.4 &  -0.4404 &   0.0444 &   2.8 &   2.7 &   4.1 &      \\
  1.402087 &   0.000066 &    35.4 &     0.6 &   0.4492 &   0.0180 &   3.2 &   4.3 &   8.7 &     1\\
  1.410727 &   0.000105 &    16.9 &     0.5 &   0.2961 &   0.0286 &   3.0 &   3.3 &   5.7 &      \\
  1.424315 &   0.000106 &    16.2 &     0.5 &   0.0104 &   0.0288 &   2.8 &   3.2 &   5.6 &      \\
  1.438538 &   0.000146 &    10.2 &     0.4 &   0.4804 &   0.0398 &   2.7 &   2.8 &   4.4 &     1\\
  1.453723 &   0.000101 &    17.9 &     0.5 &   0.3181 &   0.0276 &   2.8 &   3.3 &   5.8 &     1\\
  1.472679 &   0.000100 &    19.4 &     0.5 &  -0.0763 &   0.0273 &   2.7 &   3.2 &   5.8 &      \\
  1.479461 &   0.000155 &     9.4 &     0.4 &  -0.4500 &   0.0422 &   2.7 &   2.7 &   4.2 &      \\
  1.490630 &   0.000097 &    18.1 &     0.5 &  -0.1663 &   0.0265 &   2.7 &   3.3 &   6.0 &      \\
  1.503575 &   0.000028 &   163.0 &     1.3 &  -0.1674 &   0.0076 &   8.2 &  10.7 &  24.3 &     1\\
  1.510575 &   0.000083 &    24.8 &     0.6 &  -0.1197 &   0.0226 &   2.9 &   3.6 &   6.9 &      \\
  1.524861 &   0.000099 &    19.7 &     0.5 &   0.1852 &   0.0271 &   2.7 &   3.2 &   5.8 &      \\
  1.534509 &   0.000154 &    10.0 &     0.4 &  -0.1540 &   0.0420 &   2.7 &   2.7 &   4.2 &      \\
  1.546854 &   0.000100 &    18.3 &     0.5 &   0.4337 &   0.0273 &   2.8 &   3.2 &   5.8 &     1\\
  1.566325 &   0.000068 &    33.7 &     0.6 &  -0.0844 &   0.0185 &   3.4 &   4.2 &   8.4 &     1\\
  1.573479 &   0.000061 &    34.1 &     0.7 &  -0.4015 &   0.0166 &   3.4 &   4.6 &   9.5 &      \\
  1.582046 &   0.000068 &    35.7 &     0.6 &   0.0392 &   0.0184 &   4.4 &   4.2 &   8.5 &      \\
  1.599335 &   0.000097 &    19.3 &     0.5 &   0.1066 &   0.0264 &   2.9 &   3.3 &   6.0 &      \\
  1.604583 &   0.000177 &     8.0 &     0.4 &   0.1572 &   0.0483 &   2.7 &   2.7 &   3.9 &     1\\
  1.614715 &   0.000083 &    25.8 &     0.6 &   0.2602 &   0.0227 &   2.8 &   3.5 &   6.8 &     1\\
  1.636477 &   0.000109 &    14.0 &     0.5 &  -0.3585 &   0.0299 &   3.1 &   3.2 &   5.4 &      \\
  1.642877 &   0.000157 &     9.3 &     0.4 &   0.0686 &   0.0429 &   2.7 &   2.7 &   4.1 &      \\
  1.648307 &   0.000041 &    75.6 &     0.8 &  -0.0470 &   0.0112 &   5.2 &   6.8 &  14.6 &     1\\
  1.653515 &   0.000121 &    13.5 &     0.4 &  -0.1829 &   0.0331 &   3.1 &   3.0 &   5.0 &      \\
  1.660562 &   0.000089 &    22.1 &     0.5 &   0.3737 &   0.0243 &   2.9 &   3.5 &   6.5 &      \\
  1.668342 &   0.000067 &    33.3 &     0.6 &  -0.0822 &   0.0182 &   3.5 &   4.3 &   8.5 &      \\
  1.676784 &   0.000097 &    20.1 &     0.5 &   0.0586 &   0.0264 &   2.7 &   3.3 &   6.0 &     1\\
  1.684097 &   0.000013 &   427.2 &     1.5 &  -0.4397 &   0.0036 &  13.5 &  23.2 &  53.3 &     1\\
  1.691182 &   0.000066 &    39.0 &     0.6 &  -0.0941 &   0.0179 &   3.2 &   4.3 &   8.7 &     1\\
  1.704576 &   0.000023 &   230.0 &     1.5 &  -0.1443 &   0.0064 &   8.6 &  13.3 &  30.3 &     1\\
  1.710026 &   0.000141 &    10.8 &     0.4 &   0.1455 &   0.0384 &   2.7 &   2.8 &   4.5 &      \\
  1.725006 &   0.000152 &    10.1 &     0.4 &  -0.0918 &   0.0414 &   2.6 &   2.7 &   4.2 &      \\
  1.735050 &   0.000131 &    12.0 &     0.4 &  -0.2164 &   0.0357 &   2.9 &   2.9 &   4.7 &      \\
  1.755235 &   0.000100 &    18.7 &     0.5 &  -0.1008 &   0.0273 &   2.9 &   3.2 &   5.8 &      \\
  1.760862 &   0.000107 &    16.2 &     0.5 &  -0.2806 &   0.0292 &   2.8 &   3.2 &   5.5 &      \\
  1.801109 &   0.000114 &    14.8 &     0.5 &  -0.2217 &   0.0312 &   3.2 &   3.1 &   5.2 &      \\
  1.815474 &   0.000013 &   456.7 &     1.7 &   0.4058 &   0.0037 &  11.7 &  23.7 &  54.6 &     1\\
  1.820803 &   0.000150 &    10.0 &     0.4 &   0.0872 &   0.0408 &   2.7 &   2.7 &   4.3 &      \\
  1.833021 &   0.000068 &    32.9 &     0.6 &   0.4710 &   0.0186 &   3.4 &   4.2 &   8.4 &      \\
  1.839921 &   0.000088 &    23.1 &     0.5 &  -0.3895 &   0.0241 &   2.9 &   3.4 &   6.5 &      \\
  1.868335 &   0.000102 &    17.3 &     0.5 &  -0.2078 &   0.0277 &   3.0 &   3.3 &   5.8 &      \\
  1.874905 &   0.000105 &    16.6 &     0.5 &  -0.3295 &   0.0286 &   3.1 &   3.2 &   5.6 &      \\
  1.895515 &   0.000109 &    15.5 &     0.5 &  -0.2253 &   0.0296 &   3.3 &   3.2 &   5.5 &      \\
  1.907083 &   0.000068 &    34.7 &     0.6 &   0.2697 &   0.0185 &   3.3 &   4.2 &   8.4 &     1\\
  1.930087 &   0.000090 &    21.5 &     0.5 &  -0.1595 &   0.0246 &   3.1 &   3.4 &   6.4 &     1\\
  1.965612 &   0.000054 &    46.2 &     0.7 &   0.2402 &   0.0147 &   4.0 &   5.1 &  10.8 &     1\\
  1.975296 &   0.000090 &    21.8 &     0.5 &  -0.3080 &   0.0246 &   3.1 &   3.4 &   6.4 &      \\
  1.988790 &   0.000068 &    33.8 &     0.6 &   0.4261 &   0.0184 &   3.5 &   4.2 &   8.5 &     1\\
  1.999740 &   0.000088 &    22.3 &     0.5 &   0.1688 &   0.0239 &   3.0 &   3.4 &   6.5 &      \\
  2.004868 &   0.000036 &    91.3 &     0.9 &   0.2832 &   0.0099 &   6.5 &   7.8 &  16.9 &     1\\
  2.028060 &   0.000076 &    29.1 &     0.6 &  -0.3835 &   0.0208 &   3.4 &   3.8 &   7.5 &     1\\
  2.036727 &   0.000119 &    15.0 &     0.5 &   0.1098 &   0.0325 &   3.1 &   3.0 &   5.1 &     1\\
  2.053542 &   0.000056 &    44.7 &     0.7 &  -0.4255 &   0.0154 &   4.0 &   4.9 &  10.3 &     1\\
  2.059730 &   0.000112 &    14.9 &     0.4 &   0.3518 &   0.0305 &   3.4 &   3.2 &   5.4 &      \\
  2.086456 &   0.000156 &     9.4 &     0.4 &  -0.0760 &   0.0426 &   2.7 &   2.7 &   4.2 &      \\
  2.095714 &   0.000028 &   168.2 &     1.3 &  -0.1264 &   0.0075 &   8.0 &  10.8 &  24.5 &     1\\
  2.101907 &   0.000087 &    23.7 &     0.5 &  -0.2332 &   0.0237 &   3.2 &   3.5 &   6.6 &      \\
  2.109126 &   0.000061 &    40.4 &     0.7 &   0.3571 &   0.0168 &   3.7 &   4.6 &   9.4 &     1\\
  2.116241 &   0.000101 &    18.2 &     0.5 &  -0.2592 &   0.0275 &   2.9 &   3.3 &   5.8 &      \\
  2.144401 &   0.000100 &    17.6 &     0.5 &   0.0279 &   0.0273 &   3.0 &   3.3 &   5.8 &      \\
  2.154934 &   0.000030 &   130.3 &     1.1 &  -0.3392 &   0.0082 &   9.3 &   9.5 &  21.1 &     1\\
  2.176741 &   0.000029 &   153.8 &     1.2 &  -0.3484 &   0.0080 &   8.6 &  10.2 &  22.9 &     1\\
  2.182858 &   0.000103 &    17.2 &     0.5 &  -0.2370 &   0.0281 &   3.1 &   3.2 &   5.7 &      \\
  2.213876 &   0.000068 &    34.4 &     0.6 &   0.1517 &   0.0185 &   3.6 &   4.2 &   8.5 &     1\\
  2.227934 &   0.000029 &   137.5 &     1.1 &  -0.3339 &   0.0080 &   9.5 &   9.9 &  22.1 &     1, $2f_\mathrm{rot}$\\
  2.249873 &   0.000085 &    24.5 &     0.6 &  -0.1363 &   0.0233 &   3.2 &   3.5 &   6.7 &     1\\
  2.262957 &   0.000029 &   154.5 &     1.2 &   0.1873 &   0.0080 &   8.7 &  10.1 &  22.6 &     1\\
  2.270483 &   0.000135 &    11.3 &     0.4 &   0.0643 &   0.0368 &   2.8 &   2.9 &   4.6 &      \\
  2.289609 &   0.000103 &    17.1 &     0.5 &  -0.2617 &   0.0280 &   3.3 &   3.3 &   5.7 &      \\
  2.295325 &   0.000066 &    35.6 &     0.7 &  -0.2497 &   0.0181 &   3.9 &   4.2 &   8.6 &     1\\
  2.301730 &   0.000156 &     9.5 &     0.4 &   0.3710 &   0.0426 &   2.6 &   2.7 &   4.2 &      \\
  2.319016 &   0.000123 &    13.6 &     0.4 &  -0.4445 &   0.0334 &   3.0 &   3.0 &   5.0 &      \\
  2.326872 &   0.000084 &    24.3 &     0.6 &  -0.3121 &   0.0230 &   3.5 &   3.6 &   6.8 &     1\\
  2.337632 &   0.000161 &     8.9 &     0.4 &  -0.4458 &   0.0439 &   2.6 &   2.7 &   4.1 &     1\\
  2.345610 &   0.000120 &    13.5 &     0.4 &  -0.2043 &   0.0327 &   3.0 &   3.0 &   5.1 &     1\\
  2.369810 &   0.000161 &     8.9 &     0.4 &   0.2476 &   0.0440 &   2.6 &   2.7 &   4.1 &      \\
  2.377339 &   0.000101 &    17.9 &     0.5 &   0.0233 &   0.0276 &   3.2 &   3.3 &   5.8 &     1\\
  2.405512 &   0.000143 &    10.6 &     0.4 &  -0.2709 &   0.0390 &   2.8 &   2.8 &   4.4 &      \\
  2.428049 &   0.000067 &    33.5 &     0.6 &   0.4003 &   0.0184 &   4.4 &   4.2 &   8.5 &     1\\
  2.447994 &   0.000109 &    16.1 &     0.5 &  -0.4379 &   0.0298 &   3.1 &   3.2 &   5.5 &     1\\
  2.473258 &   0.000052 &    54.3 &     0.7 &   0.4905 &   0.0143 &   5.4 &   5.3 &  11.2 &     1\\
  2.482278 &   0.000138 &    11.1 &     0.4 &  -0.4377 &   0.0376 &   2.9 &   2.8 &   4.5 &      \\
  2.516014 &   0.000087 &    23.0 &     0.5 &  -0.3383 &   0.0238 &   3.6 &   3.5 &   6.6 &     1\\
  2.537289 &   0.000169 &     8.5 &     0.4 &  -0.4182 &   0.0460 &   2.7 &   2.8 &   4.0 &     1\\
  2.572658 &   0.000068 &    33.7 &     0.6 &   0.0960 &   0.0184 &   4.6 &   4.2 &   8.5 &     1\\
  2.582811 &   0.000114 &    15.0 &     0.5 &  -0.1934 &   0.0310 &   3.1 &   3.1 &   5.3 &      \\
  2.599784 &   0.000152 &     9.9 &     0.4 &   0.1002 &   0.0413 &   2.7 &   2.7 &   4.2 &      \\
  2.617580 &   0.000127 &    12.0 &     0.4 &  -0.3231 &   0.0346 &   3.1 &   3.0 &   4.8 &      \\
  2.648317 &   0.000164 &     8.8 &     0.4 &   0.0179 &   0.0447 &   2.7 &   2.8 &   4.1 &      \\
  2.656295 &   0.000122 &    13.0 &     0.4 &  -0.4821 &   0.0332 &   3.1 &   3.0 &   5.0 &     1\\
  2.686095 &   0.000157 &     9.4 &     0.4 &  -0.3768 &   0.0429 &   2.7 &   2.7 &   4.1 &      \\
  2.693614 &   0.000161 &     8.9 &     0.4 &   0.4999 &   0.0440 &   2.7 &   2.8 &   4.1 &      \\
  2.708373 &   0.000139 &    11.0 &     0.4 &   0.0480 &   0.0380 &   2.9 &   2.8 &   4.5 &      \\
  2.749670 &   0.000032 &   111.4 &     1.0 &   0.1421 &   0.0089 &  10.2 &   8.8 &  19.2 &     1\\
  2.808920 &   0.000127 &    12.6 &     0.4 &   0.4598 &   0.0346 &   2.9 &   3.0 &   4.8 &     1\\
  2.825828 &   0.000164 &     8.9 &     0.4 &  -0.1096 &   0.0447 &   2.7 &   2.8 &   4.1 &      \\
  2.862170 &   0.000111 &    15.0 &     0.5 &   0.1193 &   0.0303 &   2.9 &   3.1 &   5.4 &     1\\
  2.879392 &   0.000103 &    17.1 &     0.5 &   0.0951 &   0.0280 &   3.1 &   3.2 &   5.7 &     1\\
  2.930464 &   0.000123 &    13.1 &     0.4 &  -0.4801 &   0.0335 &   3.0 &   3.0 &   5.0 &     1\\
  2.969996 &   0.000068 &    33.4 &     0.6 &  -0.0488 &   0.0185 &   4.9 &   4.2 &   8.4 &     1\\
  2.977272 &   0.000117 &    14.7 &     0.5 &   0.1653 &   0.0319 &   2.9 &   3.0 &   5.1 &      \\
  2.990973 &   0.000108 &    15.6 &     0.5 &  -0.4328 &   0.0295 &   3.0 &   3.2 &   5.5 &      \\
  3.015793 &   0.000100 &    18.0 &     0.5 &  -0.0256 &   0.0273 &   3.2 &   3.2 &   5.8 &     1\\
  3.025530 &   0.000119 &    14.0 &     0.4 &  -0.4060 &   0.0325 &   2.9 &   3.0 &   5.1 &     1\\
  3.038230 &   0.000051 &    55.7 &     0.8 &  -0.3690 &   0.0139 &   6.5 &   5.4 &  11.5 &     1\\
  3.088438 &   0.000151 &     9.8 &     0.4 &   0.3462 &   0.0412 &   2.7 &   2.7 &   4.2 &     1\\
  3.097081 &   0.000251 &     5.5 &     0.3 &  -0.3872 &   0.0684 &   2.6 &   2.7 &   3.3 &     1\\
  3.102499 &   0.000059 &    43.5 &     0.7 &   0.2055 &   0.0161 &   5.7 &   4.8 &   9.8 &     1\\
  3.138301 &   0.000070 &    30.6 &     0.6 &  -0.0480 &   0.0192 &   5.1 &   4.2 &   8.1 &     1\\
  3.148135 &   0.000128 &    12.1 &     0.4 &  -0.3470 &   0.0350 &   3.0 &   3.0 &   4.8 &      \\
  3.155076 &   0.000136 &    11.9 &     0.4 &   0.0914 &   0.0372 &   2.8 &   2.9 &   4.6 &     1\\
  3.165249 &   0.000131 &    15.5 &     0.4 &   0.0683 &   0.0358 &   2.8 &   2.9 &   4.7 &      \\
  3.196142 &   0.000061 &    39.3 &     0.7 &  -0.2855 &   0.0167 &   5.9 &   4.7 &   9.4 &     1\\
  3.203072 &   0.000128 &    12.7 &     0.4 &   0.4302 &   0.0350 &   3.0 &   3.0 &   4.8 &      \\
  3.221406 &   0.000118 &    14.3 &     0.5 &  -0.1107 &   0.0321 &   3.1 &   3.0 &   5.1 &      \\
  3.254116 &   0.000100 &    18.0 &     0.5 &   0.1501 &   0.0272 &   3.4 &   3.3 &   5.9 &      \\
  3.275029 &   0.000093 &    20.3 &     0.5 &   0.3517 &   0.0255 &   3.6 &   3.4 &   6.2 &     1\\
  3.287533 &   0.000253 &     5.5 &     0.3 &   0.4839 &   0.0691 &   2.6 &   2.7 &   3.3 &     1\\
  3.318788 &   0.000032 &   120.6 &     1.0 &   0.3868 &   0.0086 &  10.9 &   9.8 &  20.0 &     1\\
  3.326295 &   0.000158 &     9.4 &     0.4 &   0.1704 &   0.0430 &   2.7 &   2.8 &   4.1 &      \\
  3.342672 &   0.000060 &    41.4 &     0.7 &  -0.2632 &   0.0165 &   5.6 &   4.9 &   9.6 &     1, $3f_\mathrm{rot}$\\
  3.357929 &   0.000152 &    10.1 &     0.4 &  -0.3047 &   0.0415 &   2.7 &   2.7 &   4.2 &     1\\
  3.369930 &   0.000136 &    11.2 &     0.4 &   0.2475 &   0.0372 &   2.9 &   2.9 &   4.6 &      \\
  3.383626 &   0.000068 &    32.1 &     0.6 &  -0.4093 &   0.0187 &   5.3 &   4.4 &   8.3 &     1\\
  3.448551 &   0.000186 &     7.4 &     0.4 &   0.2972 &   0.0508 &   2.7 &   2.8 &   3.8 &     1\\
  3.495829 &   0.000031 &   125.5 &     1.0 &  -0.4980 &   0.0085 &  10.4 &  10.2 &  20.4 &     1\\
  3.505291 &   0.000132 &    12.0 &     0.4 &   0.2798 &   0.0360 &   2.8 &   3.0 &   4.7 &     1\\
  3.554737 &   0.000154 &     9.6 &     0.4 &  -0.0788 &   0.0419 &   2.8 &   2.8 &   4.2 &      \\
  3.562600 &   0.000119 &    13.6 &     0.4 &  -0.1525 &   0.0326 &   3.1 &   3.1 &   5.1 &     1\\
  3.571243 &   0.000179 &     7.6 &     0.4 &  -0.0174 &   0.0489 &   2.7 &   2.7 &   3.8 &     1\\
  3.578247 &   0.000040 &    81.3 &     0.9 &   0.3927 &   0.0110 &   9.5 &   7.9 &  15.2 &     1\\
  3.603759 &   0.000160 &     9.1 &     0.4 &   0.1575 &   0.0436 &   2.7 &   2.7 &   4.1 &      \\
  3.611984 &   0.000109 &    15.7 &     0.5 &   0.0856 &   0.0298 &   3.3 &   3.2 &   5.5 &     1\\
  3.622813 &   0.000160 &     8.9 &     0.4 &   0.4613 &   0.0437 &   2.8 &   2.8 &   4.1 &      \\
  3.671633 &   0.000060 &    41.7 &     0.7 &   0.2519 &   0.0165 &   6.3 &   5.2 &   9.5 &     1\\
  3.724105 &   0.000093 &    20.9 &     0.5 &   0.2694 &   0.0254 &   4.1 &   3.6 &   6.2 &     1\\
  3.737811 &   0.000178 &     7.8 &     0.4 &   0.2047 &   0.0485 &   2.6 &   2.7 &   3.9 &     1\\
  3.746627 &   0.000151 &     9.5 &     0.4 &   0.0803 &   0.0412 &   2.6 &   2.8 &   4.2 &     1\\
  3.777874 &   0.000161 &     9.1 &     0.4 &   0.2032 &   0.0440 &   2.7 &   2.8 &   4.1 &      \\
  3.801212 &   0.000059 &    43.9 &     0.7 &  -0.4924 &   0.0161 &   6.4 &   5.4 &   9.8 &     1\\
  3.844671 &   0.000153 &     9.7 &     0.4 &  -0.3759 &   0.0417 &   2.7 &   2.8 &   4.2 &      \\
  3.891561 &   0.000187 &     7.6 &     0.4 &   0.0904 &   0.0510 &   2.8 &   2.8 &   3.8 &     1\\
  3.912286 &   0.000119 &    13.9 &     0.5 &  -0.1362 &   0.0324 &   3.3 &   3.2 &   5.1 &     1\\
  3.919861 &   0.000197 &     6.8 &     0.4 &   0.3739 &   0.0537 &   2.7 &   2.7 &   3.7 &     1\\
  3.967352 &   0.000085 &    23.4 &     0.6 &  -0.2119 &   0.0233 &   4.4 &   3.9 &   6.7 &     1\\
  3.982455 &   0.000236 &     5.6 &     0.4 &  -0.0206 &   0.0643 &   2.6 &   2.7 &   3.4 &     1\\
  3.989093 &   0.000145 &    10.5 &     0.4 &  -0.4542 &   0.0396 &   2.7 &   2.9 &   4.4 &      \\
  4.037990 &   0.000179 &     7.8 &     0.4 &  -0.1146 &   0.0487 &   2.5 &   2.7 &   3.8 &     1\\
  4.093517 &   0.000085 &    23.0 &     0.5 &   0.0089 &   0.0232 &   4.2 &   4.1 &   6.8 &     1\\
  4.116036 &   0.000191 &     7.9 &     0.4 &  -0.1907 &   0.0521 &   2.6 &   2.7 &   3.7 &     1\\
  4.123589 &   0.000090 &    23.2 &     0.5 &   0.0887 &   0.0246 &   4.0 &   3.9 &   6.4 &      \\
  4.204516 &   0.000099 &    18.8 &     0.5 &  -0.1637 &   0.0270 &   3.9 &   3.6 &   5.9 &     1\\
  4.229298 &   0.000155 &     9.5 &     0.4 &   0.0618 &   0.0422 &   2.6 &   2.8 &   4.2 &     1\\
  4.313953 &   0.000139 &    11.4 &     0.4 &  -0.3772 &   0.0378 &   2.6 &   3.0 &   4.5 &      \\
  4.331018 &   0.000012 &   556.3 &     1.8 &   0.3049 &   0.0033 &  34.1 &  36.1 &  62.1 &     1\\
  4.365339 &   0.000142 &    11.1 &     0.4 &  -0.1844 &   0.0387 &   2.7 &   3.0 &   4.5 &      \\
  4.416859 &   0.000150 &     9.9 &     0.4 &  -0.3079 &   0.0410 &   2.6 &   2.8 &   4.2 &     1\\
  4.430284 &   0.000119 &    13.9 &     0.4 &   0.3502 &   0.0324 &   3.1 &   3.3 &   5.1 &      \\
  4.455803 &   0.000058 &    45.0 &     0.7 &   0.1177 &   0.0158 &   6.2 &   6.2 &  10.0 &     1, $4f_\mathrm{rot}$\\
  4.518744 &   0.000122 &    13.2 &     0.4 &  -0.1685 &   0.0333 &   3.0 &   3.2 &   5.0 &      \\
  4.526613 &   0.000097 &    19.1 &     0.5 &  -0.3451 &   0.0265 &   3.7 &   3.8 &   6.0 &      \\
  4.535787 &   0.000056 &    47.3 &     0.7 &  -0.3701 &   0.0153 &   5.9 &   6.5 &  10.3 &     1\\
  4.540185 &   0.000159 &     8.9 &     0.4 &  -0.4996 &   0.0434 &   2.6 &   2.9 &   4.1 &      \\
  4.555068 &   0.000091 &    21.1 &     0.5 &   0.2178 &   0.0249 &   3.9 &   4.1 &   6.3 &     1\\
  4.565886 &   0.000131 &    12.1 &     0.4 &  -0.4432 &   0.0358 &   2.7 &   3.1 &   4.7 &      \\
  4.572236 &   0.000119 &    13.7 &     0.4 &   0.2597 &   0.0326 &   3.0 &   3.3 &   5.1 &     1\\
  4.598707 &   0.000084 &    25.4 &     0.6 &  -0.0237 &   0.0230 &   3.9 &   4.3 &   6.8 &     1\\
  4.634082 &   0.000159 &     9.4 &     0.4 &  -0.3266 &   0.0433 &   2.6 &   2.8 &   4.1 &     1\\
  4.656123 &   0.000141 &    10.5 &     0.4 &   0.2127 &   0.0384 &   2.6 &   3.0 &   4.5 &     1\\
  4.678860 &   0.000095 &    20.5 &     0.5 &   0.4466 &   0.0259 &   3.7 &   4.0 &   6.1 &     1\\
  4.708765 &   0.000160 &     9.3 &     0.4 &   0.4027 &   0.0437 &   2.8 &   2.8 &   4.1 &      \\
  4.731914 &   0.000105 &    16.4 &     0.5 &  -0.0332 &   0.0286 &   3.4 &   3.7 &   5.6 &     1\\
  4.746700 &   0.000214 &     6.2 &     0.4 &  -0.2466 &   0.0584 &   2.7 &   2.7 &   3.5 &     1\\
  4.772930 &   0.000158 &     9.8 &     0.4 &   0.0717 &   0.0432 &   2.6 &   2.8 &   4.1 &     1\\
  4.800770 &   0.000139 &    11.0 &     0.4 &   0.2749 &   0.0380 &   2.6 &   3.1 &   4.5 &     1\\
  4.810913 &   0.000163 &     8.8 &     0.4 &   0.0801 &   0.0445 &   2.7 &   2.9 &   4.1 &      \\
  4.853467 &   0.000153 &     9.9 &     0.4 &   0.1046 &   0.0418 &   2.7 &   2.9 &   4.2 &      \\
  4.867541 &   0.000088 &    22.3 &     0.5 &   0.1266 &   0.0240 &   3.6 &   4.3 &   6.5 &     1\\
  4.921036 &   0.000069 &    31.6 &     0.6 &  -0.3456 &   0.0189 &   4.6 &   5.7 &   8.2 &     1\\
  4.929812 &   0.000150 &     9.9 &     0.4 &  -0.3600 &   0.0409 &   2.6 &   2.9 &   4.3 &      \\
  4.938572 &   0.000072 &    30.5 &     0.6 &   0.3753 &   0.0196 &   4.6 &   5.5 &   7.9 &     1\\
  4.995854 &   0.000141 &    10.5 &     0.4 &  -0.1811 &   0.0386 &   2.8 &   3.1 &   4.5 &      \\
  5.004592 &   0.000043 &    72.1 &     0.8 &  -0.4147 &   0.0117 &   8.2 &  10.2 &  14.0 &     1\\
  5.012089 &   0.000060 &    41.6 &     0.7 &   0.0488 &   0.0163 &   5.9 &   7.0 &   9.7 &      \\
  5.018511 &   0.000119 &    14.1 &     0.5 &  -0.1806 &   0.0324 &   2.9 &   3.4 &   5.1 &      \\
  5.025314 &   0.000045 &    65.3 &     0.8 &  -0.0129 &   0.0123 &   8.2 &   9.6 &  13.2 &     1\\
  5.035087 &   0.000134 &    11.5 &     0.4 &  -0.2094 &   0.0365 &   2.7 &   3.2 &   4.7 &      \\
  5.059220 &   0.000089 &    21.9 &     0.5 &   0.2833 &   0.0242 &   3.8 &   4.4 &   6.5 &     1\\
  5.079514 &   0.000152 &     9.7 &     0.4 &   0.2235 &   0.0415 &   2.6 &   2.9 &   4.2 &     1\\
  5.176795 &   0.000100 &    18.0 &     0.5 &  -0.0246 &   0.0272 &   3.4 &   4.0 &   5.8 &     1\\
  5.204155 &   0.000158 &     9.3 &     0.4 &   0.4320 &   0.0432 &   2.7 &   2.9 &   4.2 &      \\
  5.214582 &   0.000194 &     7.2 &     0.4 &   0.0676 &   0.0528 &   2.7 &   2.8 &   3.7 &     1\\
  5.221908 &   0.000103 &    17.1 &     0.5 &   0.0410 &   0.0280 &   3.4 &   3.9 &   5.7 &      \\
  5.239917 &   0.000132 &    11.8 &     0.4 &  -0.4920 &   0.0359 &   2.8 &   3.2 &   4.7 &      \\
  5.250504 &   0.000163 &     8.9 &     0.4 &  -0.2291 &   0.0445 &   2.7 &   2.9 &   4.1 &      \\
  5.281941 &   0.000127 &    13.1 &     0.4 &   0.2144 &   0.0346 &   2.9 &   3.3 &   4.8 &      \\
  5.310661 &   0.000128 &    12.1 &     0.4 &   0.0660 &   0.0349 &   3.0 &   3.3 &   4.8 &      \\
  5.329144 &   0.000126 &    13.3 &     0.4 &   0.4657 &   0.0343 &   2.9 &   3.3 &   4.9 &     1\\
  5.344036 &   0.000163 &     8.9 &     0.4 &  -0.4587 &   0.0446 &   2.7 &   2.9 &   4.1 &     1\\
  5.410919 &   0.000114 &    14.7 &     0.5 &  -0.3762 &   0.0311 &   3.2 &   3.6 &   5.3 &     1\\
  5.440438 &   0.000139 &    11.0 &     0.4 &   0.3339 &   0.0380 &   3.0 &   3.1 &   4.5 &      \\
  5.487003 &   0.000100 &    17.7 &     0.5 &   0.4337 &   0.0274 &   3.7 &   4.1 &   5.8 &     1\\
  5.499241 &   0.000036 &    95.1 &     0.9 &   0.2946 &   0.0099 &  11.3 &  14.2 &  17.0 &     1\\
  5.512107 &   0.000142 &    10.7 &     0.4 &   0.1805 &   0.0386 &   3.1 &   3.2 &   4.5 &     1\\
  5.533914 &   0.000127 &    12.7 &     0.4 &  -0.1343 &   0.0346 &   3.0 &   3.3 &   4.8 &      \\
  5.559664 &   0.000101 &    17.9 &     0.5 &   0.3300 &   0.0276 &   3.6 &   4.1 &   5.8 &     1\\
  5.569338 &   0.000164 &     8.7 &     0.4 &  -0.4035 &   0.0447 &   2.7 &   2.9 &   4.1 &     $5f_\mathrm{rot}$\\
  5.673346 &   0.000051 &    55.5 &     0.8 &  -0.0727 &   0.0139 &   9.2 &   9.7 &  11.5 &     1\\
  5.692809 &   0.000163 &     9.1 &     0.4 &   0.1939 &   0.0444 &   2.8 &   3.0 &   4.1 &      \\
  5.841228 &   0.000068 &    33.0 &     0.6 &   0.2816 &   0.0185 &   6.5 &   6.7 &   8.4 &     1\\
  5.962605 &   0.000119 &    13.6 &     0.4 &  -0.0465 &   0.0326 &   3.7 &   3.7 &   5.1 &     1\\
  5.979591 &   0.000216 &     6.2 &     0.4 &  -0.3948 &   0.0588 &   2.6 &   2.8 &   3.5 &     1\\
  5.995156 &   0.000108 &    16.2 &     0.5 &   0.1730 &   0.0294 &   4.1 &   4.1 &   5.5 &     1\\
  6.014117 &   0.000132 &    11.7 &     0.4 &  -0.1517 &   0.0361 &   3.5 &   3.4 &   4.7 &      \\
  6.022806 &   0.000157 &     9.4 &     0.4 &  -0.0775 &   0.0427 &   2.9 &   3.1 &   4.2 &      \\
  6.051632 &   0.000150 &     9.9 &     0.4 &   0.3180 &   0.0409 &   3.0 &   3.1 &   4.2 &     1\\
  6.104056 &   0.000090 &    21.7 &     0.5 &   0.1408 &   0.0244 &   4.9 &   5.0 &   6.4 &     1\\
  6.137888 &   0.000160 &     9.1 &     0.4 &  -0.2583 &   0.0436 &   2.9 &   3.1 &   4.1 &     1\\
  6.184881 &   0.000106 &    16.1 &     0.5 &  -0.1463 &   0.0290 &   4.1 &   4.3 &   5.6 &     1\\
  6.214041 &   0.000215 &     5.9 &     0.4 &   0.0360 &   0.0586 &   2.8 &   2.8 &   3.5 &     1\\
  6.231460 &   0.000131 &    11.7 &     0.4 &  -0.4820 &   0.0358 &   3.4 &   3.5 &   4.7 &      \\
  6.312680 &   0.000165 &     8.6 &     0.4 &  -0.0300 &   0.0450 &   3.0 &   3.1 &   4.0 &     1\\
  6.328771 &   0.000099 &    18.7 &     0.5 &  -0.4110 &   0.0271 &   4.3 &   4.6 &   5.8 &     1\\
  6.357912 &   0.000043 &    71.0 &     0.8 &  -0.3885 &   0.0117 &  12.5 &  13.3 &  13.9 &     1\\
  6.424794 &   0.000126 &    12.8 &     0.4 &  -0.2383 &   0.0345 &   3.6 &   3.6 &   4.8 &      \\
  6.558738 &   0.000097 &    19.1 &     0.5 &  -0.4807 &   0.0264 &   5.3 &   5.2 &   6.0 &     1\\
  6.568077 &   0.000155 &     9.5 &     0.4 &  -0.1485 &   0.0423 &   3.2 &   3.3 &   4.2 &      \\
  6.683187 &   0.000121 &    13.4 &     0.4 &   0.4685 &   0.0329 &   3.8 &   3.9 &   5.0 &     1, $6f_\mathrm{rot}$\\
  6.708996 &   0.000140 &    10.6 &     0.4 &  -0.2087 &   0.0383 &   4.0 &   3.6 &   4.5 &      \\
  6.743097 &   0.000131 &    11.8 &     0.4 &  -0.1795 &   0.0358 &   3.6 &   3.7 &   4.7 &     1\\
  6.763042 &   0.000104 &    16.8 &     0.5 &   0.4728 &   0.0283 &   4.7 &   4.6 &   5.7 &     1\\
  6.790082 &   0.000089 &    22.0 &     0.5 &  -0.4884 &   0.0242 &   5.1 &   5.5 &   6.5 &      \\
  6.873458 &   0.000199 &     6.8 &     0.4 &  -0.3051 &   0.0543 &   2.8 &   2.9 &   3.6 &     1\\
  6.911532 &   0.000212 &     6.3 &     0.4 &   0.1067 &   0.0578 &   2.9 &   2.9 &   3.5 &     1\\
  6.961982 &   0.000159 &     9.1 &     0.4 &   0.0419 &   0.0433 &   3.4 &   3.3 &   4.1 &     1\\
  7.156625 &   0.000132 &    11.7 &     0.4 &   0.1171 &   0.0359 &   4.1 &   3.8 &   4.7 &     1\\
  7.198978 &   0.000097 &    19.6 &     0.5 &   0.4574 &   0.0264 &   5.3 &   5.2 &   6.0 &     1\\
  7.211881 &   0.000140 &    10.8 &     0.4 &  -0.0817 &   0.0382 &   4.0 &   3.7 &   4.5 &      \\
  7.302012 &   0.000130 &    12.1 &     0.4 &   0.1517 &   0.0354 &   4.1 &   3.9 &   4.8 &     1\\
  7.313601 &   0.000223 &     5.9 &     0.4 &   0.1754 &   0.0608 &   2.8 &   2.9 &   3.4 &     1\\
  7.334501 &   0.000151 &     9.7 &     0.4 &   0.1424 &   0.0413 &   3.8 &   3.5 &   4.2 &      \\
  7.394462 &   0.000068 &    33.0 &     0.6 &  -0.1994 &   0.0186 &   8.8 &   8.2 &   8.4 &     1\\
  7.481286 &   0.000127 &    12.6 &     0.4 &   0.4712 &   0.0346 &   4.1 &   4.1 &   4.8 &     1\\
  7.649805 &   0.000108 &    15.7 &     0.5 &   0.3498 &   0.0294 &   4.8 &   5.0 &   5.5 &     1\\
  7.658273 &   0.000329 &     3.8 &     0.3 &  -0.0494 &   0.0897 &   2.6 &   2.7 &   2.9 &     1\\
  7.797888 &   0.000142 &    10.9 &     0.4 &   0.0105 &   0.0387 &   3.9 &   4.0 &   4.5 &     $7f_\mathrm{rot}$\\
  7.825444 &   0.000203 &     6.7 &     0.4 &  -0.4112 &   0.0553 &   3.1 &   3.1 &   3.6 &     1\\
  7.854532 &   0.000155 &     9.6 &     0.4 &   0.3025 &   0.0423 &   3.7 &   3.7 &   4.2 &     1\\
  7.903092 &   0.000175 &     7.9 &     0.4 &   0.4861 &   0.0478 &   3.3 &   3.4 &   3.9 &     1\\
  8.011471 &   0.000184 &     7.5 &     0.4 &  -0.4904 &   0.0503 &   3.4 &   3.3 &   3.8 &     1\\
  8.109830 &   0.000155 &     9.5 &     0.4 &   0.2905 &   0.0422 &   3.8 &   3.8 &   4.2 &     1\\
  8.212746 &   0.000188 &     7.4 &     0.4 &   0.4659 &   0.0513 &   3.3 &   3.3 &   3.8 &     1\\
  8.758742 &   0.000175 &     8.1 &     0.4 &  -0.1966 &   0.0476 &   3.9 &   3.7 &   3.9 &     1\\
  8.786500 &   0.000199 &     6.8 &     0.4 &  -0.2775 &   0.0543 &   3.4 &   3.3 &   3.7 &     1\\
  9.176342 &   0.000231 &     5.7 &     0.4 &  -0.2763 &   0.0629 &   3.0 &   3.1 &   3.4 &     1\\
  9.237725 &   0.000124 &    12.7 &     0.4 &   0.2322 &   0.0338 &   5.5 &   5.3 &   4.9 &     1\\
  9.289781 &   0.000215 &     6.2 &     0.4 &   0.3784 &   0.0587 &   3.1 &   3.3 &   3.5 &     1\\
  9.328187 &   0.000102 &    17.2 &     0.5 &  -0.1407 &   0.0279 &   7.0 &   6.8 &   5.8 &     1\\
  9.426554 &   0.000142 &    10.8 &     0.4 &  -0.1972 &   0.0387 &   4.7 &   4.7 &   4.5 &      \\
  9.611562 &   0.000199 &     6.8 &     0.4 &   0.2532 &   0.0543 &   3.4 &   3.5 &   3.7 &     1\\
  9.758696 &   0.000159 &     9.1 &     0.4 &  -0.3708 &   0.0434 &   4.4 &   4.4 &   4.1 &     1\\
 10.024449 &   0.000121 &    13.3 &     0.4 &   0.3130 &   0.0330 &   6.3 &   6.0 &   5.0 &     1\\
 10.689655 &   0.000114 &    14.8 &     0.5 &   0.1147 &   0.0310 &   7.0 &   6.9 &   5.3 &     1\\
 10.890249 &   0.000054 &    49.0 &     0.7 &  -0.0045 &   0.0148 &  17.0 &  19.5 &  10.7 &     1\\
 10.976678 &   0.000158 &     9.3 &     0.4 &   0.4653 &   0.0431 &   5.0 &   4.8 &   4.1 &     1\\
 11.140515 &   0.000225 &     5.7 &     0.4 &   0.4347 &   0.0613 &   3.5 &   3.4 &   3.4 &     1\\
 11.300558 &   0.000203 &     6.6 &     0.4 &  -0.0809 &   0.0553 &   3.8 &   3.9 &   3.6 &     1\\
 12.575224 &   0.000134 &    11.5 &     0.4 &   0.2842 &   0.0365 &   6.6 &   6.5 &   4.7 &     1\\
 12.597091 &   0.000215 &     6.2 &     0.4 &  -0.4142 &   0.0588 &   4.0 &   3.9 &   3.5 &     1\\
 13.078566 &   0.000183 &     7.7 &     0.4 &   0.4403 &   0.0499 &   4.6 &   4.8 &   3.8 &     1\\
 13.151984 &   0.000138 &    11.1 &     0.4 &  -0.1711 &   0.0376 &   6.3 &   6.6 &   4.5 &      \\
 13.537170 &   0.000193 &     7.0 &     0.4 &  -0.4695 &   0.0528 &   4.3 &   4.7 &   3.7 &     1\\
 15.022126 &   0.000222 &     5.9 &     0.4 &   0.3025 &   0.0607 &   4.9 &   4.5 &   3.4 &     1\\
 15.043053 &   0.000182 &     7.4 &     0.4 &  -0.4717 &   0.0498 &   5.4 &   5.5 &   3.8 &     1\\
 15.220953 &   0.000099 &    18.3 &     0.5 &   0.2037 &   0.0271 &   9.5 &  10.9 &   5.9 &     1\\
 15.307708 &   0.000159 &     9.6 &     0.4 &  -0.1230 &   0.0433 &   6.0 &   6.4 &   4.1 &     1\\
 15.408226 &   0.000219 &     6.1 &     0.4 &  -0.0691 &   0.0599 &   4.8 &   4.7 &   3.5 &     1\\
 15.521060 &   0.000206 &     6.4 &     0.4 &   0.2494 &   0.0561 &   5.0 &   5.0 &   3.6 &     1\\
 15.793377 &   0.000292 &     4.3 &     0.3 &   0.4660 &   0.0798 &   3.8 &   3.7 &   3.1 &     1\\
 16.806802 &   0.000149 &    10.1 &     0.4 &   0.2465 &   0.0407 &   7.3 &   7.6 &   4.3 &      \\
 16.823663 &   0.000129 &    12.2 &     0.4 &  -0.2181 &   0.0352 &   8.1 &   9.0 &   4.8 &     1\\
 18.589675 &   0.000218 &     6.1 &     0.4 &  -0.4386 &   0.0593 &   5.4 &   5.2 &   3.5 &     1\\
 18.821437 &   0.000172 &     8.2 &     0.4 &  -0.4517 &   0.0469 &   7.0 &   6.7 &   4.0 &     1\\
\end{longtable}

\end{appendix}

\end{document}